\documentclass[fleqn,usenatbib]{mnras}

\usepackage{newtxtext,newtxmath}
\usepackage{multirow}
\usepackage[caption=false]{subfig}

\usepackage[T1]{fontenc}

\DeclareRobustCommand{\VAN}[3]{#2}
\let\VANthebibliography\thebibliography
\def\thebibliography{\DeclareRobustCommand{\VAN}[3]{##3}\VANthebibliography}

\usepackage{graphicx}	
\usepackage{amsmath}	

\title[Understanding the blazar sequence]{Understanding the phenomenological and intrinsic blazar sequence using a simple scaling model}

\author[Z.-J. Wan et al.]{
Zhu-Jian Wan,$^{1}$
Rui Xue,$^{1}$\thanks{E-mail: ruixue@zjnu.edu.cn}
Ze-Rui Wang,$^{2}$\thanks{E-mail: zerui\_wang62@163.com}
Hu-Bing Xiao,$^{3}$\thanks{E-mail: hubing.xiao@shnu.edu.cn}
and Jun-Hui Fan$^{4,5,6}$\thanks{E-mail: fjh@gzhu.edu.cn}
\\
$^{1}$Department of Physics, Zhejiang Normal University, Jinhua 321004, China\\
$^{2}$College of Physics and Electronic Engineering, Qilu Normal University, Jinan 250200, China\\
$^{3}$Shanghai Key Lab for Astrophysics, Shanghai Normal University, Shanghai 200234, People's Republic of China\\
$^{4}$Center for Astrophysics, Guangzhou University, Guangzhou 510006, People's Republic of China\\
$^{5}$Key Laboratory for Astronomical Observation and Technology of Guangzhou, Guangzhou, 510006, People's Republic of China\\
$^{6}$Astronomy Science and Technology Research Laboratory of Department of Education of Guangdong Province, Guangzhou, 510006, People's Republic of China
}

\date{Accepted XXX. Received YYY; in original form ZZZ}

\pubyear{2024}

\begin{document}
\label{firstpage}
\pagerange{\pageref{firstpage}--\pageref{lastpage}}
\maketitle

\begin{abstract}
The blazar sequence, including negative correlations between radiative luminosity $L_{\rm rad}$ and synchrotron peak frequency $\nu$, and between Compton dominance $Y$ and $\nu$, is widely adopted as a phenomenological description of spectral energy distributions (SEDs) of blazars, although its underlying cause is hotly debated. In particular, these correlations turn positive after correcting Doppler boosting effect. In this work, we revisit the phenomenological and intrinsic blazar sequence with three samples, which are historical sample (SEDs are built with historical data), quasi-simultaneous sample (SEDs are built with quasi-simultaneous data) and Doppler factor corrected sample (a sample with available Doppler factors), selected from literature. We find that phenomenological blazar sequence holds in historical sample, but does not exist in quasi-simultaneous sample, and intrinsic correlation between $L_{\rm rad}$ and $\nu$ becomes positive in Doppler factor corrected sample. We also analyze if the blazar sequence still exists in subclasses of blazars, i.e., flat-spectrum radio quasars and BL Lacertae objects, with different values of $Y$. To interpret these correlations, we apply a simple scaling model, in which physical parameters of the dissipation region are connected to the location of the dissipation region. We find that the model generated results are highly sensitive to the chosen ranges and distributions of physical parameters. Therefore, we suggest that even though the simple scaling model can reproduce the blazar sequence under specific conditions that have been fine-tuned, such results may not have universal applicability. Further consideration of a more realistic emission model is expected.
\end{abstract}

\begin{keywords}
radiation mechanisms: non-thermal -- galaxies: active -- galaxies: jets.
\end{keywords}



\section{Introduction}\label{intro}
Blazars are the most extreme active galactic nuclei (AGNs) with relativistic jets aligned to our line of sight \citep{1995PASP..107..803U}. The spectral energy distribution (SED) of blazars consists of two broad non-thermal emission components in the $\log\nu -\log\nu L_{\nu}$ diagram. The low-energy component, peaking between the infrared to X-ray bands, is believed to originate from the synchrotron emission of relativistic electrons within the jet. The high-energy component, peaking in the $\gamma$-ray band, is believed to originate from the inverse Compton (IC) emission of the same electron population that up-scatter, either the synchrotron photons emitted by the same population of relativistic electrons \citep[synchrotron self-Compton, SSC;][]{1981ApJ...243..700K} or external photons \citep[external Compton, EC;][]{1989ApJ...340..181G,1992A&A...256L..27D,1993ApJ...416..458D,2002ApJ...577...78S} from an accretion disc, a broad-line region (BLR), or a dusty torus (DT). Based on the equivalent width (EW) of the emission lines, blazars are divided into flat-spectrum radio quasars (FSRQs) with strong broad emission lines (EW$>5$ \AA), and BL Lacertae objects (BL Lacs) with absent or weak emission lines \citep[EW$\leqslant5$ \AA;][]{1996MNRAS.281..425M, 2004MNRAS.351...83L, 2014MNRAS.441.3375X}. \cite{2011MNRAS.414.2674G} introduce a physical distinction between these two subclasses based on the BLR luminosity $L_{\rm BLR}$ in units of the Eddington luminosity $L_{\rm Edd}$ of the central supermassive black hole (SMBH). FSRQs are objects whose $L_{\rm BLR}\gtrsim 5 \times 10^{-4}L_{\rm Edd}$, and BL Lacs are the others. In addition to the FSRQs and BL Lacs classification, \cite{2010ApJ...710.1271A} divide blazars into high-synchrotron-peaked (HSP) blazars if their synchrotron peak $\nu_{\rm s}^{\rm p}>10^{15}~\rm Hz$, intermediate-synchrotron-peaked (ISP) blazars if their synchrotron peak $10^{14}~\rm Hz<\nu_{\rm s}^{\rm p}<10^{15}~\rm Hz$ and low-synchrotron-peaked (LSP) blazars if their synchrotron peak $\nu_{\rm s}^{\rm p}<10^{14}~\rm Hz$.

Based on a sample of 126 blazars, \cite{1998MNRAS.299..433F} identified that their observed radiative luminosity, and the Compton dominance (the ratio of fluxes or luminosities of high- and low-energy components) are both anti-correlated to the observed synchrotron peak frequency. These two negative correlations are referred to as the well-known ``blazar sequence''. The first physical explanation is proposed by \cite{1998MNRAS.301..451G}. They suggested that if the intrinsic jet power is connected to $L_{\rm BLR}$, one would expect the synchrotron peak frequency shifts from high frequency to low frequency by increasing the intrinsic jet power and increasing the radiative cooling dominated by the EC scattering. If this interpretation is correct, it provides a powerful tool for understanding the evolution of blazars' SEDs, similar to the evolution of stars on the Hertzsprung--Russel diagram. Theoretical models suggest that the jet power is generated from accretion and the extraction of rotational energy or angular momentum from the disc/black hole \citep{1977MNRAS.179..433B, 1982MNRAS.199..883B}, therefore the jet properties and emissions are linked to the black hole mass and accretion rate. \cite{2008MNRAS.387.1669G} further highlight that the black hole mass and accretion rate are connected to the jet power and the shape of SED, providing a framework that could help explain the existence of both low-luminosity `blue' quasars and low-luminosity `red' quasars. The importance of the accretion rate is also pointed out by \cite{2011ApJ...740...98M}. They find that AGNs in the blazar sequence diagram is mainly composed of two groups, the primary difference between these groups being the accretion rate. The group with a high accretion rate consists mostly of LSP blazars and FR II radio galaxies, while the group with a low accretion rate corresponds to the majority of BL Lacs and FR I radio galaxies. In addition, many other studies endeavor to come up with physical interpretations of the blazar sequence. For instance, \cite{2002ApJ...564...86B} propose a simple analytical model, suggesting that the reduction of gas and dust in external environment causes the evolutionary sequence from FSRQ to BL Lacs-LSPs, and then to BL Lacs-HSPs. \cite{2010ApJ...723..417B} propose that the blazar sequence can be well understood in the case of multiple IC scatterings, and the energy density of relativistic electrons is the main factor influencing the formation of the blazar sequence. \cite{2013ApJ...763..134F} construct a simple model to explain the blazar sequence in the 2LAC clean sample, suggesting that the difference between sources is associated with the magnetic field of the jet's dissipation region, the energy density of the external photon field, and the jet's viewing angle. Note that all sources in the modeling are assumed to have the same bulk Lorentz factor. \cite{2013MNRAS.436..304P} propose a relativistic continuous jet model, which suggests that the blazar sequence could be interpreted if the radius of the transition region and bulk Lorentz factor of the jet increase with the jet power. According to their interpretation, FSRQs have large bulk Lorentz factors, and scatter external cosmic microwave background photons at large distances, resulting in larger Compton dominance and lower IC peak frequency. On the other hand, BL Lacs, which have lower jet powers, possess stronger magnetic fields in the bright synchrotron transition region compared to high-power FSRQs, leading to higher peak frequency synchrotron and SSC emissions.

Since the discovery of the blazar sequence, its underlying cause and if it is a result of selection effect have been widely discussed, but even so it is still frequently adopted as a description of the blazar population. With the improvement of detector sensitivity and the expansion of samples, the blazar sequence has been confirmed and developed \citep{2008MNRAS.387.1669G, 2011ApJ...735..108C, 2013ApJ...763..134F, 2015MNRAS.450.3568X, 2016Galax...4...36G, 2016ApJS..224...26M, 2017MNRAS.469..255G, 2022MNRAS.509.4620S, 2023ApJ...951..133K, 2023ApJ...949...52O}. \cite{2013ApJ...763..134F} suggests that the redshift selection effect cannot explain the blazar sequence. In \cite{2017MNRAS.469..255G}, a revised version of the blazar sequence, named the ``\textit{Fermi} blazar sequence'', is propose by using a clean sample of 747 blazars (299 BL Lacs and 448 FSRQs) from the third catalogue of AGN detected by \textit{Fermi}-LAT \citep{2015ApJ...810...14A}. In this large blazar sample, the existence of the blazar sequence is confirmed. However it is also pointed out that when considering BL Lacs and FSRQs separately, the sequence still holds for BL Lacs but not for FSRQs. On the other hand, the blazar sequence is questioned by the discovery of ``outlier'' blazars (e.g., luminous HSP blazars) and the evidence of selection effects \citep{2006A&A...445..441N, 2012MNRAS.422L..48P, 2012A&A...541A.160G, 2012MNRAS.420.2899G, 2015MNRAS.450.2404G, 2016A&A...587A...8R, 2017A&A...606A..68C, 2021MNRAS.505.4726K}. 

In the original physical interpretation \citep{1998MNRAS.301..451G}, radiative cooling is suggested as the main cause of the two negative correlations in the phenomenological blazar sequence. If so, these two negative correlations should firstly hold in the comoving frame, therefore it has to be assumed that all blazars have similar values of the Doppler factor, or the Doppler factor is not related to other physical parameters \citep{2017MNRAS.469..255G, 2022Galax..10...35P}. However, some studies argue that the phenomenological blazar sequence is an artefact of the Doppler boosting, since these two correlations with Doppler-corrected values turn positive or disappear \citep{2008MNRAS.391..967L, 2008A&A...488..867N, 2009RAA.....9..168W, 2015MNRAS.451.2750X, 2016RAA....16..173F, 2017ApJ...835L..38F, 2021ApJ...906..108C, 2022ApJ...925..120Y}. It poses a challenge to the physical understanding of the blazar sequence. 

In this work, we are motivated to revisit the phenomenological and intrinsic blazar sequence of samples in the literature, and propose a theoretical interpretation to these correlations. In previous studies, the blazar sequence is studied with historical SEDs. However, since blazars are highly variable objects, we also check if the blazar sequence holds for the quasi-simultaneous SEDs. This paper is organized as follows.  In Sect.~\ref{revisit}, we revisit the blazar sequence with three samples. In Sect.~\ref{theory}, we propose a theoretical consideration to understand the phenomenological and intrinsic blazar sequence. Finally, we end with the conclusion in Sect.~\ref{DC}. Throughout the paper, the cosmological parameters $H_{0}=69.6\ \rm km\ s^{-1}Mpc^{-1}$, $\Omega_{0}=0.29$, and $\Omega_{\Lambda}$= 0.71 are adopted \citep{2014ApJ...794..135B}. 

\section{Revisiting the blazar sequence with the $Fermi$ blazar}\label{revisit}
The original blazar sequence is a phenomenological description of the blazars' SED, including the negative correlations between the synchrotron peak frequency and the radiative luminosity, and between the synchrotron peak frequency and Compton dominance. In this section, we revisit these two correlations both in the observers' frame and in the comoving frame with different samples. 

When studying the correlation between the radiative luminosity and the synchrotron peak frequency in the blazar sequence, radio luminosity was originally used \citep[e.g.,][]{1998MNRAS.299..433F} since it is less variable than other bands luminosity \citep{2012A&A...541A.160G} and can serve as a good indicator of the radiative luminosity \citep{2017Ap&SS.362..191W}. However, it should be noted that using radio luminosity may introduces two sources of bias: (i) radio emission is generally believed to originate from the extended jet because of the synchrotron self-absorption, while emissions from other bands come from the inner jet, indicating that they originate from different locations of the jet and have different physical origins \citep{2009MNRAS.399.2041G, 2010MNRAS.402..497G}; (ii) blazars are highly variable objects, with significant variations in both the luminosities and the peak frequencies of the two humps \citep[e.g.,][]{2004A&A...413..489M, 2011ApJ...729....2A,2017ApJS..229...21X,2020ApJS..247...49X}. Although radio luminosity may serve as a long-term proxy for radiative luminosity, there is no quantity available to serve as a long-term proxy for the synchrotron peak frequency. 
Therefore, we suggest using the integrated full-band jet luminosity as a proxy for radiative luminosity instead of radio luminosity, which can mitigate the first bias. The bias due to variability is difficult to avoid, but this approach can at least ensure that the full-band integrated luminosity and the synchrotron peak frequency originate from the same snapshot of SED, making their physical connection more intimate. Thanks to the abundant multi-wavelength observations at present, a more accurate estimate of the radiative luminosity can be obtained by fitting the full-band SED. To investigate the impact of variability on correlations of the blazar sequence, it would be intriguing to test if non-simultaneous and quasi-simultaneous SEDs exhibit similar behavior.

\subsection{Samples}\label{sample}
This work focuses on the blazar sequence and uses $Fermi$ blazars as samples, since $\gamma$-ray luminosity is a good indicator of radiative luminosity \citep{2017Ap&SS.362..191W} and is crucial for calculating Compton dominance. We study three samples selected from the literature using linear correlations:
\begin{enumerate}
    \item The ``historical sample'', compiled by \cite{2022ApJS..262...18Y, 2023SCPMA..6649511Y} from the Fourth \textit{Fermi}-LAT 12-year Source catalog \citep{2022ApJS..260...53A}, contains 750 FSRQs and 844 BL Lacs with measured redshifts. The historical SEDs of all blazars are constructed, and a parabolic equation is used to fit them\footnote{The low-energy component is fitted in \cite{2022ApJS..262...18Y} and the high-energy component is fitted in \cite{2023SCPMA..6649511Y}.}, enabling the derivation of Compton dominance $Y$, as well as synchrotron peak frequency $\nu^{\rm obs}$ and radiative luminosity $L_{\rm rad}^{\rm obs}$ in observers' frame.
    \item \cite{2016MNRAS.463.3038X} collect quasi-simultaneous multi-wavelength data of 279 blazars from the second \textit{Fermi}-LAT AGN catalog \citep{2011ApJ...743..171A}. In their sample, 81 FSRQs and 28 BL Lacs have the full-waveband quasi-simultaneous SEDs, which are also fitted using a parabolic equation. This sample is referred to as the ``quasi-simultaneous sample'' hereafter.
    \item By cross-matching the \textit{Fermi}-4FGL blazars collected in \cite{2022ApJS..262...18Y, 2023SCPMA..6649511Y} with the sample of \cite{2018ApJ...866..137L}, \cite{2022ApJ...925..120Y} find 180 blazars, including  129 FSRQs and 51 BL Lacs, with available Doppler factors $\delta_{\rm D}$ and measured redshifts. Hereafter, this sample is referred to as the ``$\delta_{\rm D}$-corrected sample''. Due to the beaming effect, the synchrotron peak frequency $\nu$ and radiative luminosity $L_{\rm rad}$ in the comoving frame can be obtained through $\nu\simeq \nu^{\rm obs} \delta_{\rm D}^{-1}$ and $L_{\rm rad}\simeq L_{\rm rad}^{\rm obs}\delta_{\rm D}^{-4}$, respectively. Please note that this $\delta_{\rm D}$-corrected sample is highly biased, as there are only three sources with $Y<1$. This implies that there are clear selection biases when using this sample for correlation studies. If a more complete sample can be collected in the future, the corresponding correlation results could be more reliable and meaningful.
\end{enumerate}
In the above samples, the synchrotron peak frequency, the radiative luminosity, and Compton dominance are all obtained by fitting SEDs with parabolic equations. However, one may argue that fitting a non-standard parabola SED with the parabolic equation may introduce bias in correlation studies. In \cite{2016MNRAS.463.3038X}, they respectively fit apparent asymmetrical SEDs of some specific blazars using the parabolic equation and the cubic polynomial, and find that the difference in obtained luminosities is insignificant. In \cite{2022ApJS..262...18Y}, they compare the synchrotron peak frequency derived from parabolic equation with those derived from the cubic polynomial of \cite{2010ApJ...716...30A}. They find that the difference of the average peak frequency obtained by different fitting methods is minor. Therefore, we suggest that it is reasonable to use the parameters obtained by the parabolic equation to revisit the blazar sequence.

In the following, we use historical and quasi-simultaneous samples to revisit the phenomenological blazar sequence and compare their correlation results. The $\delta_{\rm D}$-corrected sample enables us to study the intrinsic blazar sequence. It should be noted that the obtained Doppler factor is based on the variability brightness temperature of radio observations \citep{2018ApJ...866..137L}. Although \cite{2019ApJ...870...28F} finds that Doppler factors in parsec-scale jets obtained from the conical jet model \citep{1979ApJ...232...34B} are basically in agreement with those obtained from radio observations \citep{2009A&A...494..527H}, the agreement is within quite significant errors. It should be noted that obtaining accurate $\delta_{\rm D}$ for the dominant dissipation region is incredibly difficult, and even the values measured by radio observations at different times for the same source can vary greatly. For example, for the same sample of blazars, values of $\delta_{\rm D}$ obtained by \cite{2018ApJ...866..137L} has a large discrepancy with those obtained by \cite{2009A&A...494..527H} and \cite{2017MNRAS.466.4625L}. Therefore, using $\delta_{\rm D}$ measured by radio observation to study the intrinsic blazar sequence will inevitably introduce a large uncertainty and potential observational biases.

\begin{figure}
\centering
\includegraphics[scale=0.5]{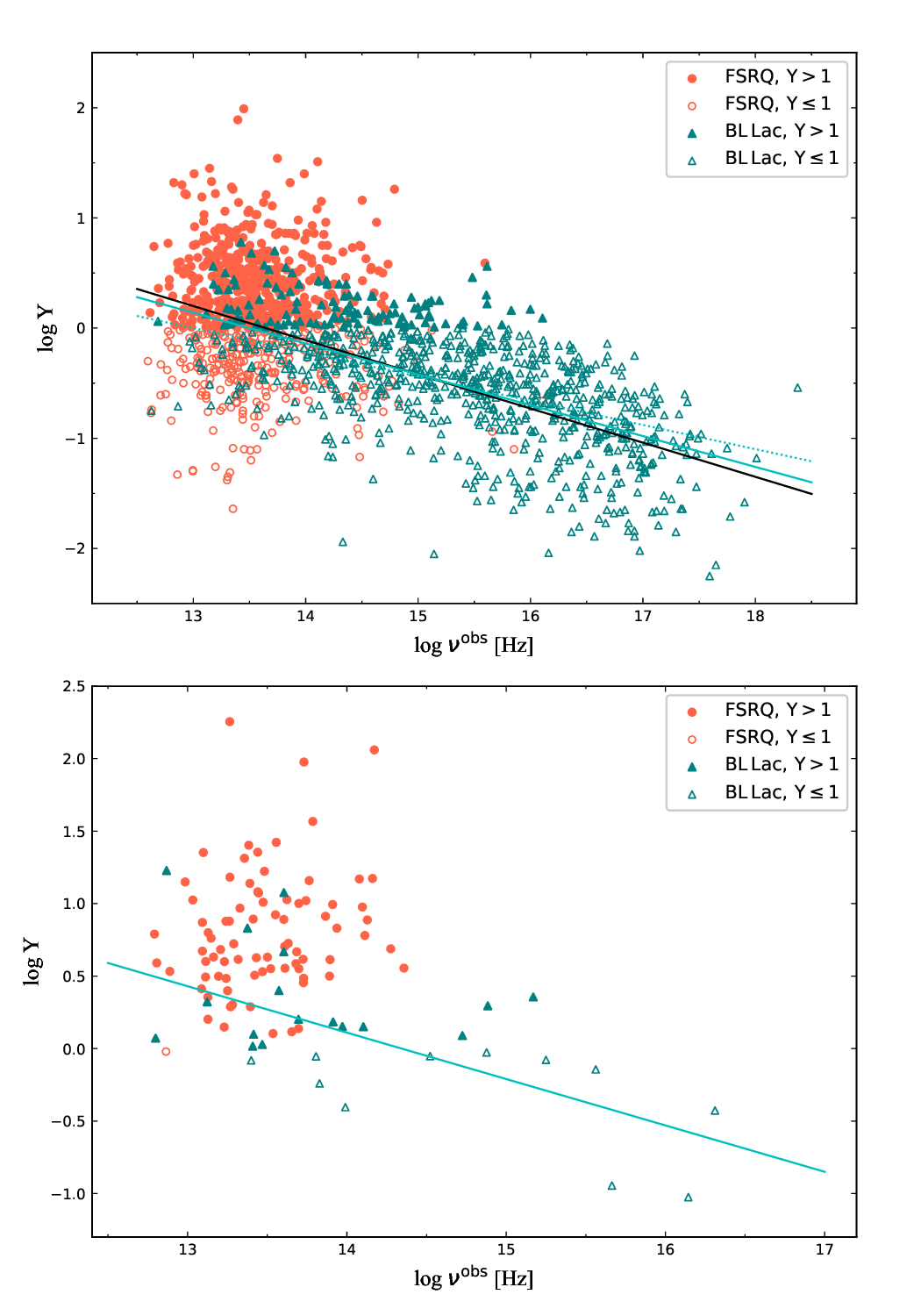}
\caption{The correlations between $\log (Y)$ and $\log (\nu^{\rm obs})$ in the historical sample (upper panel), and in the quasi-simultaneous sample (lower panel).
As explained in the inset legends, the red symbol represents the FSRQs and the teal symbol represents the BL Lacs, the full symbol represents the source with $Y>1$ and the empty symbol represents the source with $Y\le 1$.
If a significant correlation is found ($\tau>0.1,~p<0.01$) for a specific (sub-) sample statistically, the best linear fitting equation is also shown. 
The solid black line represents the best-fitting equation for the whole sample. 
The solid red line represents the best-fitting equation for the whole population of FSRQs, 
the dashed red line represents the best-fitting equation for the population of FSRQs with $Y>1$, 
and the dotted red line represents the best-fitting equation for the population of FSRQs with $Y\le 1$.
The solid teal line represents the best-fitting equation for the whole population of BL Lacs, 
the dashed teal line represents the best-fitting equation for the population of BL Lacs with $Y>1$, 
and the dotted teal line represents the best-fitting equation for the population of BL Lacs with $Y\le 1$.}
\label{correlations_Y}
\end{figure}

\subsection{Correlation Results}\label{CR}
We analyze the phenomenological blazar sequence in both the historical and quasi-simultaneous samples, and the intrinsic blazar sequence in $\delta_{\rm D}$-corrected sample. We do not study the phenomenological blazar sequence in $\delta_{\rm D}$-corrected sample further, since the $\delta_{\rm D}$-corrected sample is a subsample of historical sample. The theoretical analysis proposed in Sect.~\ref{theory} indicates that FSRQs and BL Lacs with $Y \le 1$ and $Y>1$ may have different correlations and slopes, so we further divide three samples collected in this work using the dividing line $Y=1$. Table~\ref{table1} shows the statistical correlation results, including slopes of the best linear fitting equations (denoted by $s$ hereafter). Based on the Spearman rank correlation test, our results are described as follows: correlation coefficients between $0.10$ and $0.29$ indicate a weak correlation, coefficients between $0.30$ and $0.49$ represent a moderate correlation, and coefficients between $0.50$ and $1$ represent a strong correlation. If the chance probabilities are $p > 0.01$, it is suggested that the correlation is not established statistically, which implies that we cannot rule out the possibility that the correlation result is due to chance factors \citep{cohen2013applied}.

For the correlation between $\log(Y)$ and $\log(\nu^{\rm obs})$ in the observers' frame, we find a strong negative correlation ($\tau=-0.59$) for all blazars in historical sample, consistent with previous studies \citep{1998MNRAS.299..433F,2011ApJ...735..108C}. While, for the quasi-simultaneous sample, only a weak negative correlation ($\tau=-0.18$) is found with $p=0.06$, which might be attributed to the smaller sample size. When considering FSRQs separately, the correlation disappears in historical sample, while the correlation result for the quasi-simultaneous sample remains basically unchanged, since this sample is dominated by FSRQs. This result is consistent with that of \cite{2013ApJ...763..134F} using data from the second catalogue of AGN detected by \textit{Fermi}-LAT. For BL Lacs, we find strong negative correlations in both the historical ($\tau=-0.64$) and quasi-simultaneous sample ($\tau=-0.53$), which are consistent with \cite{2013ApJ...763..134F}. Strong negative correlations are also found for BL Lacs with $Y\le 1$ in the both the historical ($\tau=-0.58$) and quasi-simultaneous ($\tau=-0.54$) samples. However it should be noted that a large chance probability ($p=0.09$) is obtained for the quasi-simultaneous sample, which may caused by the small sample size. Also it is interesting that no significant correlation is found for BL Lacs with $Y>1$ in either the historical or quasi-simultaneous sample. The above correlation results are displayed in Figure \ref{correlations_Y}.

\begin{figure}
\centering
\includegraphics[scale=0.5]{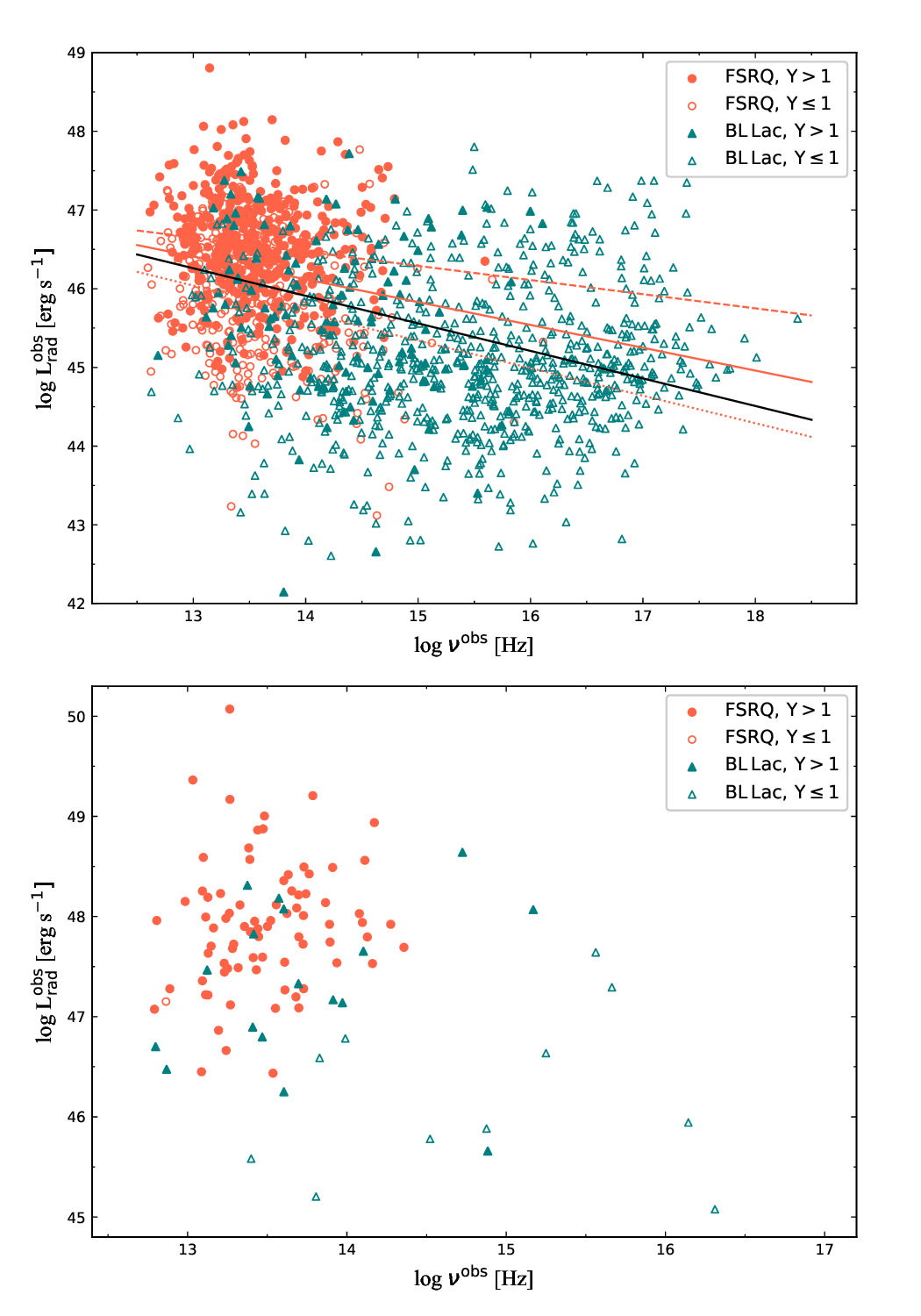}
\caption{The correlations between $\log (L_{\rm rad}^{\rm obs})$ and $\log (\nu^{\rm obs})$ in the historical sample (upper panel), and in the quasi-simultaneous sample (lower panel).
The meanings of symbols and line styles are the same as in Figure \ref{correlations_Y}. Similarly, the best linear fitting equation is only shown when a significant correlation is found.}
\label{correlations_L}
\end{figure}

\begin{figure}
\centering
\includegraphics[scale=0.5]{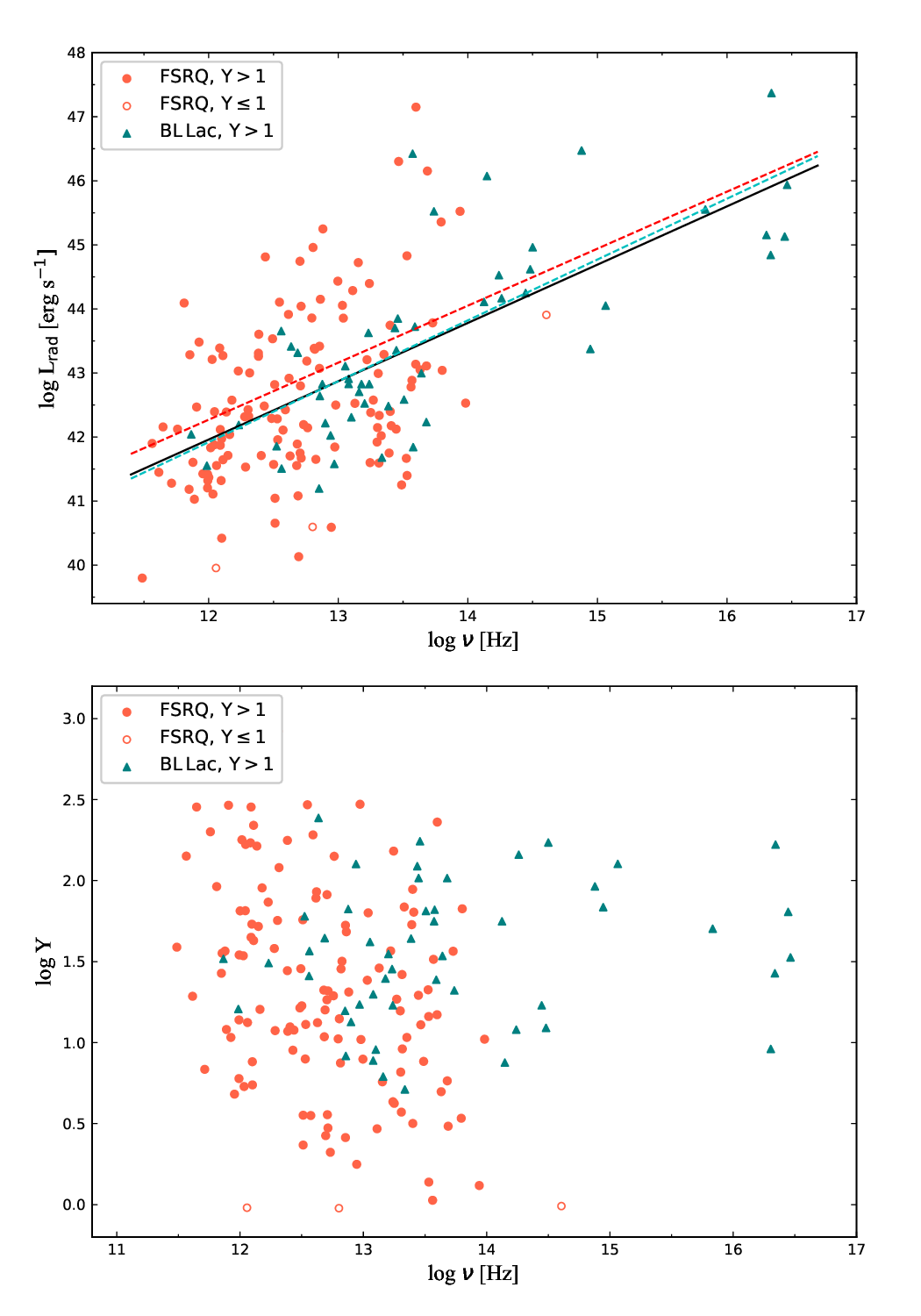}
\caption{The upper panel shows the correlations between $\log (L_{\rm rad})$ and $\log (\nu)$, and the lower panel shows the correlations between $\log (Y)$ and $\log (\nu)$ in the $\delta_{\rm D}$-corrected sample. The meanings of symbols and line styles are the same as in Figure \ref{correlations_Y}. Similarly, the best linear fitting equation is only shown when a significant correlation is found.}
\label{correlations_corrected sample}
\end{figure}

The correlations between $\log(L^{\rm obs}_{\rm rad})$ and $\log(\nu^{\rm obs})$ in the observers' frame for the historical and quasi-simultaneous samples are shown in Figure \ref{correlations_L}. A moderate negative correlation ($\tau=-0.48$) is found in historical sample, which is consistent with many previous studies \citep{1998MNRAS.299..433F,2011ApJ...735..108C,2013ApJ...763..134F,2015MNRAS.450.3568X,2016ApJS..226...20F,2017MNRAS.469..255G}, but no correlation ($\tau=-0.05$) is found in quasi-simultaneous sample. When focusing on FSRQs, weak negative correlations are found both in historical ($\tau=-0.14$) and quasi-simultaneous samples ($\tau=-0.2$; also a large chance probability $p=0.08$ is obtained for quasi-simultaneous sample). For BL Lacs in historical sample, no correlation is found for the whole sample and for subsample with $Y\le 1$. For BL Lacs with  $Y > 1$, only a weak correlation with a large chance probability ($\tau=-0.17,~p=0.07$) is found. For BL Lacs in quasi-simultaneous sample, weak correlations are found, but the chance probabilities are significant high ($p>0.3$), suggesting that these correlations are not significant. Among the above results of the whole sample in the historical sample, our main focus is on the correlation between the full-band radiative luminosity and the peak frequency. In contrast, \cite{2022ApJS..262...18Y} focuses on the correlations between the peak frequency and radio, optical, X-ray, and $\gamma$-ray emissions, respectively. Given that most FSRQs have $Y>1$, which means the $\gamma$-ray luminosity dominates the full-band radiative luminosity, our results are basically consistent with \cite{2022ApJS..262...18Y}. However, for BL Lacs, the band that dominates the full-band radiative luminosity is uncertain. Therefore, it is difficult to directly compare our results with \cite{2022ApJS..262...18Y}, which focuses on the correlation of a specific band. Our correlation study suggests that no correlation is found for the whole sample of BL Lacs. As can be seen from the upper panel of Figure \ref{correlations_L}, no correlation found for the whole sample of BL Lacs may be due to the discovery of many high-luminosity HSPs and low-luminosity LSPs in the Fourth Fermi-LAT 12-year Source catalog. It is worth noting that our findings differ from those obtained using data from the third catalogue of AGN detected by \textit{Fermi}-LAT (3LAC), where correlations are found in BL Lacs instead of FSRQs \citep{2017MNRAS.469..255G}. In addition, \cite{2016ApJS..226...20F} find marginal correlations in BL Lacs-HSPs ($\tau=0.1,~p=3.6\times 10^{-3}$) instead of FSRQs or BL Lacs-LSPs, and \cite{2016ApJS..224...26M} find strong correlations without particular separation between FSRQs and BL Lacs.

Based on the results above, there are similarities and differences in correlation studies of the phenomenological blazar sequence between the historical and quasi-simultaneous samples. Some of these differences may be attributed to the fact that the quasi-simultaneous sample contains different states of SEDs \citep{2016MNRAS.463.3038X}, while others may be due to the small size of the quasi-simultaneous sample, resulting in $p > 0.01$. Moreover, for the quasi-simultaneous SED given by \citep{2016MNRAS.463.3038X}, the observation time for radio, optical, and X-ray is within one week, while the $\gamma$-ray data has been integrated over several months. However, variabilities on the scales of days \citep[e.g.,][]{2012ApJ...756...13B}, hours \citep[e.g.,][]{2015ApJ...807...79H}, and even minutes \citep[e.g.,][]{2008MNRAS.384L..19B} have been widely observed in different energy bands of blazars. Therefore, using the quasi-simultaneous data collected by \citep{2016MNRAS.463.3038X} will introduce potential bias. It is interesting that if the correlation holds both for the historical and quasi-simultaneous samples, their $s$ values are also quite close. In the subsequent theoretical interpretation (see Sect.~\ref{theory}), we choose to rely on the results of the historical sample. A larger quasi-simultaneous sample with high-quality and same state (high or low) SEDs is needed to check these results in detail.

\begin{table*}
\centering
\caption{Correlation results of historical, quasi-simultaneous and $\delta_{\rm D}$-corrected samples. Columns from left to right: (1): the sample studied in correlation analysis; (2): the number of blazars in the sample; (3) and (6) are the Spearman test correlation coefficients; (4) and (7) are the chance probabilities; (5) and (8) are the slopes of the best linear fitting equations.}
\label{table1}
\begin{tabular}{cccccccc} 
\hline\hline
 \multirow{2}{*}{ in the observers' frame }                         & \multirow{2}{*}{N} & \multicolumn{3}{c}{$\log (Y)~ {\rm vs.}~ \log (\nu^{{\rm obs}})$}                                & \multicolumn{3}{c}{$ \log (L_{{\rm rad}}^{{\rm obs}})~ {\rm vs.}~ \log (\nu^{{\rm {obs}}})$}              \\ 
\cline{3-8}
    &        &$\tau$      & $p$   & $s$   & $\tau$    & $p$  & $s$ \\ 
    (1)&(2)&(3)&(4)&(5)&(6)&(7)&(8)\\
\hline
ALL, historical sample       & 1594               & -0.59   & $5.5\times10^{-152}$         & $-0.31 \pm  0.009$ & -0.48 & $2.0\times10^{-91}$ & $-0.35 \pm 0.02$  \\
ALL, quasi-simultaneous sample     & 109                & -0.18   & 0.06  & $ -0.34\pm 0.07$    & -0.05 & 0.63               & $-0.38 \pm 0.12 $  \\ 
\hline
FSRQs, historical sample     &                    &                 &                                    &                 &               &                            &                \\ 
\hline
ALL                         & 750                & -0.002   & 0.95                        & $-0.04\pm 0.04$    & -0.14 & $7.8\times10^{-5}$ & $-0.29\pm 0.06$   \\
$Y\le 1$             & 288                & -0.005  & 0.93                        & $-0.02 \pm 0.03 $  & -0.20 & $5.0\times10^{-4}$ & $-0.35\pm 0.08 $  \\
$Y>1$            & 462                & $5.28\times 10^{-4}$ & 0.98                        & $8.05\times 10^{-4} \pm 0.04$ & -0.14 & 0.002              & $-0.18\pm 0.07$   \\ 
\hline
FSRQs, quasi-simultaneous sample   &                    &                 &                                    &                 &               &                            &                \\ 
\hline
ALL                         & 81                 & -0.22   & 0.05                        & $-0.25\pm 0.13$   & -0.20   & 0.08                & $-0.29\pm 0.20$    \\ 
\hline
BL Lacs, historical sample   &                    &                 &                                    &                 &               &                            &                \\ 
\hline
ALL                         & 844                & -0.64   & $9.6\times10^{-97}$         & $-0.28\pm 0.01$    & -0.04 & 0.19                & $-0.02\pm 0.03$   \\
$Y\le 1$             & 726                & --0.58   & $3.3\times10^{-65}$         & $-0.22\pm 0.01$    & 0.07   & 0.07                & $0.08\pm 0.04$    \\
$Y>1$            & 118                & -0.19   & 0.04                        & $-0.06\pm 0.02$    & -0.17 & 0.07                & $-0.21\pm 0.14$   \\ 
\hline
BL Lacs, quasi-simultaneous sample &                    &                 &                                    &                 &               &                            &                \\ 
\hline
ALL                         & 28                 & -0.53   & 0.004 & $-0.32\pm 0.08$    & -0.18 & 0.37                & $-0.22\pm 0.19$   \\
$Y\le 1$             & 11                 & -0.54   & 0.09                        & $-0.20 \pm 0.10$    & 0.21   & 0.54                & $0.13\pm 0.27$    \\
$Y>1$            & 17                 & -0.03   & 0.90                        & $-0.14\pm 0.14$    & 0.15   & 0.56                & $0.20\pm 0.31$    \\
\hline
in the comoving frame& & \multicolumn{3}{c}{$\log (Y)~ {\rm vs.}~ \log (\nu)$}& \multicolumn{3}{c}{$ \log (L_{{\rm rad}})~ {\rm vs.}~ \log (\nu)$}  \\ 
\hline
ALL, $\delta_{\rm D}$-corrected sample & 180                                                             & -0.10 & 0.17         & $-0.03 \pm 0.05$  & 0.54 & $8.8\times10^{-15}$  & $0.91 \pm 0.09$                       \\ 
FSRQs, $Y>1$     & 126                                                                &   -0.23            &     0.12          &    -$0.24 \pm 0.08$            &      0.40       &   $4.4\times10^{-6}$    &      $0.89 \pm 0.17$                              \\ 

BL Lacs, $Y>1$    &   51                                                              &    0.16           &    0.15          &    $0.06 \pm 0.05$            &      0.76       &  $8.8\times10^{-15}$     &   $0.95 \pm 0.12$                                 \\ 

\hline\hline
\end{tabular}
\end{table*}

We study the intrinsic blazar sequence using the $\delta_{\rm D}$-corrected sample, and the correlation results are presented in Figure \ref{correlations_corrected sample}. For the correlation between $\log (Y)$ and $\log (\nu)$, weak correlations are found, but the obtained large chance probabilities ($p>0.1$) suggesting that these correlations might not be significant. In contrast, a strong positive correlation ($\tau=0.54$) is found for the correlation between $\log (L_{\rm rad})$ and $\log (\nu)$. Furthermore, when we study FSRQs and BL Lacs (dominated by ones with $Y>1$), respectively, similar correlation results and slopes ($s\sim1$) are derived as well. The obtained positive correlations are consistent with those obtained by some previous works \citep{2008A&A...488..867N, 2009RAA.....9..168W, 2016RAA....16..173F, 2021ApJ...906..108C}, and the obtained slopes are consistent with our previous studies \citep{2017ApJ...835L..38F, 2022ApJ...925..120Y}. It is noteworthy that $\delta_{\rm D}$ used by \cite{2021ApJ...906..108C} are derived from the conventional one-zone leptonic model \citep{2018ApJS..235...39C}. However, it is unfortunate that the slope of the best linear fitting equation is omitted in their study.

\section{Theoretical analysis}\label{theory}
\subsection{Simple scaling model}\label{SDM}
In the study of blazars, many phenomenological models have been proposed. However, most of these models contain numerous free parameters, even the simplest conventional one-zone model has at least seven free parameters that are coupled to each other. This may not be inappropriate for studying the global properties of blazars, such as the blazar sequence. In radio and X-ray observations, bright knots have been observed along the jet of blazars and radio galaxies \citep[e.g.,][]{2011ApJ...729...26M, 2016A&A...595A..54M}. In other words, significant dissipation regions can potentially appear at any location along the jet \citep{2007MNRAS.380....2W, 2009MNRAS.397..985G, 2023MNRAS.526.5054L}. This kind of framework that considers the dominant dissipation region appearing at different positions has been applied by \cite{2018MNRAS.473.4107P} and \cite{2022PhRvD.105b3005W} to explain the orphan flares of blazars. Following this phenomenological framework, we assume that one single spherical dissipation region appearing at different position of the jet dominates the whole jet emission, which is certainly an over-simplified assumption \citep[for a multi-zone view, please see][]{2023MNRAS.526.5054L}. This dominant dissipation region is composed of a plasma of charged particles in a uniform magnetic field $B$ with radius $R$ and moving with bulk Lorentz factor $\Gamma=(1-\beta^2)^{-1/2}$, where $\beta c$ is the jet speed, along the jet, at a viewing angle $\theta^{\rm obs}$ with respect to observers' line of sight. By assuming $\theta^{\rm obs} \lesssim 1/\Gamma$ for blazars, we have $\delta_{\rm D} \approx \Gamma$. Let us envision that the dominant dissipation regions of all blazars can be placed within one jet. In this way, these dominant dissipation regions are distributed at different positions within this jet. Based on some reasonable assumptions that usually applied in the continuous jet model \citep[e.g.,][]{1979ApJ...232...34B}, physical parameters of these dissipation regions are connected with each other. Radio observations have found that AGNs jets have an approximate conical \citep{2007ApJ...668L..27K, 2011A&A...532A..38S} or parabolic structure \citep{2013ApJ...775..118N, 2017ApJ...834...65A, 2018NatAs...2..472G, 2018MNRAS.475..368H}. If assuming that the dissipation region occupies the entire jet cross section, we have $R\propto r^\alpha$, where $r$ is the distance between SMBH and the dissipation region. More specifically, $\alpha=1$ represents the jet has a conical structure, and $0<\alpha<1$ means the jet has a parabolic structure. For the magnetic field $B$, we have $B\propto R^{-1}\propto r^{-\alpha}$ by assuming the magnetic power is conserved in the jet \citep{1979ApJ...232...34B}, which is a crucial assumption for reproducing the flat radio spectrum \citep{2012MNRAS.423..756P}. In addition, if considering the jet's continuous acceleration or deceleration as suggested by phenomenological model, magnetohydrodynamic simulation, and observation \citep{1979ApJ...232...34B, 2006ApJ...641..103H, 2013MNRAS.429.1189P, 2014A&A...570A...7M, 2024MNRAS.52710262R}, we assume a simplified acceleration/deceleration profile, i.e., $\delta_{\rm D} \propto r^{x}$, where $x>0$ for an accelerating jet, $x<0$ for a decelerating jet, and $x=0$ for a constant-speed jet or $\delta_{\rm D}$ does not related to other parameters. In this way, power-law scaling connections of $R$, $B$, $\delta_{\rm D}$ and $r$ can be established. With this simple scaling model, the potential physical origins of the statistical correlation of the blazar sequence can be investigated. Of course, one should bear in mind that the jet each dissipation region located in is actually different, therefore the normalization of each power-law correlations is still a free parameter. In this section, parameters with superscript ``obs'' are measured in observers' frame, whereas the parameters without the superscript are measured in the comoving frame, unless specified otherwise.

Assuming that accelerated relativistic electrons are injected in the dissipation region with a specific spectral shape $\Omega(\gamma)$ at a constant rate given by $\dot{Q}(\gamma)=\dot{Q}_0\Omega(\gamma), \gamma_{\rm min}<\gamma<\gamma_{\rm max},$ where $\gamma_{\rm min}$ is the the minimum electron Lorentz factor, $\gamma_{\rm max}$ is the maximum electron Lorentz factor, and $\dot{Q}_0$ is the normalization in units of $\rm cm^{-3}~s^{-1}$. By giving an electron injection luminosity $L_{\rm inj}$, $\dot{Q}_0$ can be calculated by 
\begin{equation}\label{Q0}
\dot{Q}_0=\frac{L_{\rm inj}}{V_{\rm b}mc^2\int_{\gamma_{\rm min}}^{\gamma_{\rm max}} \gamma \Omega(\gamma)d\gamma},
\end{equation}
where $m$ is the rest mass of an electron, $c$ is the speed of light, and $V_{\rm b}$ is the volume of the dissipation region. The radiative luminosity (i.e., the bolometric luminosity) in the comoving frame can be calculated as 
\begin{equation}\label{Lrad1}
L_{\rm rad}=V_{\rm b}mc^2\int_{\gamma_{\rm min}}^{\gamma_{\rm max}} f(\gamma)\gamma \dot{Q}(\gamma)d\gamma,
\end{equation}
where $f(\gamma)=\rm min\{\frac{\it{t}_{\rm dyn}}{\it{t}_{\rm cool}}, 1\}$ is the cooling efficiency of electrons. More specifically, $t_{\rm dyn}=\eta R/c$ with $\eta>1$ is the dynamical timescale\footnote{Undoubtedly, this is clearly an oversimplification, since the modeling of the particle escape requires a detailed specification of the geometry, the boundary conditions, the magnetic-field configuration. The particle's dynamical timescale is likely to be energy dependent of the energy-dependent diffusion. In phenomenological blazar model, the form applied here is to represent the average dynamical timescale \citep[e.g.,][]{2013LNP...873.....G}, when considering the escaping electrons at different position of the spherical blob.}, and $t_{\rm cool}=3mc/\left[4\gamma \sigma_{\rm T}U_{\rm B}(1+Y) \right]$ is the radiative cooling timescale, where $\sigma_{\rm T}$ is the Thomson scattering cross-section, $U_{\rm B}=B^2/(8\pi)$ is the energy density of the magnetic field, and $Y$ is the Compton dominance. In this work, we aim to study if radiative cooling is responsible for the blazar sequence, so it is more appropriate to use the full-band radiative luminosity $L_{\rm rad}$.

Substituting Eq.~(\ref{Q0}) into Eq.~(\ref{Lrad1}), we have
\begin{equation}\label{Lrad2}
\begin{split}
L_{\rm rad}&=\frac{L_{\rm inj}}{\int_{\gamma_{\rm min}}^{\gamma_{\rm max}} \gamma \Omega(\gamma)d\gamma}\int_{\gamma_{\rm min}}^{\gamma_{\rm max}}f(\gamma)\gamma \Omega(\gamma)d\gamma \\
&=\frac{L_{\rm inj}}{\int_{\gamma_{\rm min}}^{\gamma_{\rm max}} \gamma \Omega(\gamma)d\gamma}
\bigg\{\int_{\gamma_{\rm c}}^{\gamma_{\rm max}} \gamma \Omega(\gamma)d\gamma+
\frac{f(\gamma)}{\gamma}\int_{\gamma_{\rm min}}^{\gamma_{\rm c}}\gamma^2 \Omega(\gamma)d\gamma \bigg\} \\
&\propto L_{\rm inj}(\gamma>\gamma_{\rm c})+L_{\rm inj}(\gamma<\gamma_{\rm c})B^2R(1+Y)
\end{split}
\end{equation}
where $\gamma_{\rm c}=\frac{3mc^2}{4\sigma_{\rm T}RU_{\rm B}(1+Y)}$ represents the cooling break at which $t_{\rm dyn}=t_{\rm cool}$.

Following the physical interpretation of \cite{1998MNRAS.301..451G}, if we believe that the peak of the low-energy component is caused by radiative cooling, the peak frequency in the comoving frame can be estimated using the monochromatic approximation \citep{1998ApJ...509..608T} 
\begin{equation}\label{nu_c}
\begin{split}
\nu&\approx 3.7\times10^6 \gamma_{\rm c}^2B \\
&\propto B^{-3}R^{-2}(1+Y)^{-2}
\end{split} .
\end{equation}
From Eqs.~(\ref{Lrad2}) and (\ref{nu_c}), it can be seen that expressions of $L_{\rm rad}$, $\nu$ and $Y$ (its expression will be given later) consist of several physical parameters coupled together, including $R$, $B$, and $\delta_{\rm D}$. Considering connections of $R$, $B$, $\delta_{\rm D}$ and $r$ in the framework of the simple scaling model, $L_{\rm rad}$ becomes
\begin{equation}\label{L_rad3}
L_{\rm rad} \propto L_{\rm inj}(\gamma>\gamma_{\rm c})+L_{\rm inj}(\gamma<\gamma_{\rm c})(1+Y)r^{-\alpha},
\end{equation}
and $\nu$ becomes
\begin{equation}\label{nu_peak}
\nu\propto r^{\alpha}(1+Y)^{-2}.
\end{equation}

For FSRQs, the Compton dominance $Y$ can be evaluated as 
\begin{equation}\label{Y_F}
\begin{split}
Y&\simeq \frac{\delta_{\rm D}^2U_{\rm ext}}{U_{\rm B}} \\
&\propto \left\{
\begin{array}{rl}
r^{2\alpha+2x}, r<r_{\rm ext} \\ 
r^{2\alpha+2x-n}, r>r_{\rm ext}
\end{array} \right.,
\end{split}
\end{equation}
where $U_{\rm ext}$ is the energy density of the external photon field in the AGN frame, and $r_{\rm ext}$ represents the characteristic distance of external photon fields that is found to scale with the disk luminosity $L_{\rm D}$ as $r_{\rm ext}\propto {L_{\rm D}}^{1/2}$ through reverberation mapping \citep{2006ApJ...639...46S, 2011A&A...536A..78K, 2013ApJ...767..149B, 2014A&A...561L...8P}. When the dissipation region is located within $r_{\rm ext}$, $U_{\rm ext}=L_{\rm D}/(4\pi r_{\rm ext}^2c)$ would be a constant value \citep{2020NatCo..11.5475H}; when the dissipation region is located beyond $r_{\rm ext}$, $U_{\rm ext}$ decreases with the index $n$ \citep{2009ApJ...704...38S, 2012ApJ...754..114H}. From Eqs.~(\ref{nu_peak}) and (\ref{Y_F}), the non-linear correlation between the synchrotron peak frequency and the Compton dominance in logarithmic space can be obtained
\begin{equation}\label{FYnonjet}
\begin{split}
\nu \propto \left\{
\begin{array}{rl}
Y^{\frac{\alpha}{2\alpha+2x}}(1+Y)^{-2}, r<r_{\rm ext} \\
Y^{\frac{\alpha}{2\alpha+2x-n}}(1+Y)^{-2}, r>r_{\rm ext} \\
\end{array} \right.
\end{split}
\end{equation}
in the comoving frame, and 
\begin{equation}\label{FYnonobs}
\begin{split}
\nu^{\rm obs} \propto \left\{
\begin{array}{rl}
Y^{\frac{1}{2}}(1+Y)^{-2}, r<r_{\rm ext} \\
Y^{\frac{\alpha+x}{2\alpha+2x-n}}(1+Y)^{-2}, r>r_{\rm ext} \\
\end{array} \right.
\end{split}
\end{equation}
in the observers' frame. If considering $Y\ll1$ and $Y\gg1$\footnote{In the following, whenever symbols `$Y\ll1$' or `$Y\gg1$' appear, it implies that the actual correlation is non-linear in logarithmic space when $Y\sim1$.},  above correlations become linear in logarithmic space, i.e.,
\begin{equation}\label{FY<1}
\begin{split}
Y\propto \left\{
\begin{array}{rl}
{\nu}^{\frac{2\alpha+2x}{\alpha}}, r<r_{\rm ext} \\
{\nu}^{\frac{2\alpha+2x-n}{\alpha}}, r>r_{\rm ext}
\end{array} \right., Y\ll1\\
Y\propto \left\{
\begin{array}{rl}
{\nu}^{\frac{2\alpha+2x}{-3\alpha-4x}}, r<r_{\rm ext} \\
{\nu}^{\frac{2\alpha+2x-n}{2n-3\alpha-4x}}, r>r_{\rm ext}
\end{array} \right., Y\gg1
\end{split}
\end{equation}
in the comoving frame, and 
\begin{equation}\label{Eq12}
\begin{split}
Y\propto \left\{
\begin{array}{rl}
{\nu^{\rm obs}}^2, r<r_{\rm ext} \\
{\nu^{\rm obs}}^{\frac{2\alpha+2x-n}{\alpha+x}}, r>r_{\rm ext}
\end{array} \right., Y\ll1\\
Y\propto \left\{
\begin{array}{rl}
{\nu^{\rm obs}}^{-2/3}, r<r_{\rm ext} \\
{\nu^{\rm obs}}^{\frac{2\alpha+2x-n}{2n-3\alpha-3x}}, r>r_{\rm ext}
\end{array} \right., Y\gg1
\end{split}
\end{equation}
in the observers' frame. It should be noted that due to the actual correlation being non-linear, when $Y\sim1$, under certain combinations of $\alpha$, $n$, and $x$, the true slope may have significant errors compared to those when $Y\ll1$ and $Y\gg1$.
From Eq.~(\ref{L_rad3}), it can be seen that if the first term on the right is dominant, the correlation between $\nu$ and $L_{\rm rad}$ depends on if there is a correlation between $\nu$ and $L_{\rm inj}$. Assuming that the first term on the right is not dominant and values of $L_{\rm inj}$ do not have a large dispersion, we have
\begin{equation}\label{FLC}
\begin{split}
L_{\rm rad}\propto\left\{
\begin{array}{rl}
&{\nu}^{-1}, \it{Y}\ll1 \\
&{\nu}^{\frac{2x+\alpha}{-3\alpha-4x}}, \it{Y}\gg1, r<r_{\rm ext} \\
&{\nu}^{\frac{2x+\alpha-n}{2n-3\alpha-4x}}, \it{Y}\gg1, r>r_{\rm ext}
\end{array} \right.
\end{split}
\end{equation}
in the comoving frame, and 
\begin{equation}\label{FLO}
\begin{split}
L_{\rm rad}^{\rm obs}\propto\left\{
\begin{array}{rl}
&{\nu^{\rm obs}}^{\frac{4x-\alpha}{x+\alpha}}, \it{Y}\ll1 \\
&{\nu^{\rm obs}}^{\frac{6x+\alpha}{-3\alpha-3x}}, \it{Y}\gg1, r<r_{\rm ext} \\
&{\nu^{\rm obs}}^{\frac{6x+\alpha-n}{2n-3\alpha-3x}}, \it{Y}\gg1, r>r_{\rm ext}
\end{array} \right.
\end{split}
\end{equation}
in the observers' frame. Note that the above analysis is based on the assumption that radiative cooling induces the spectrum break. However cooling induced change of spectral index is rarely discovered in observations \citep{2022MNRAS.514.3074B}. Therefore, it is necessary to re-study these two correlations for $\nu<\nu_{\rm c}$. In this scenario, electrons are cooled in slow cooling regime, so the break of the electron energy distribution corresponding to the peak frequency of the low-energy component might be ascribed to multiple acceleration processes \citep{1978MNRAS.182..443B, 2023arXiv231113873T}. In the diffusive shock acceleration, the electron Lorentz factor can be evaluated as $\gamma \propto BR$ \citep{1977ICRC...11..132A, 1978MNRAS.182..147B, 1978MNRAS.182..443B, 1978ApJ...221L..29B, 1988A&A...201..177K, 2007Ap&SS.309..119R}, therefore $\nu\propto B^3r^2\propto r^{-\alpha}$. Then, for the correlation between the synchrotron peak frequency and the Compton dominance, we have
\begin{equation}\label{Eq16}
Y\propto \left\{
\begin{array}{rl}
{\nu}^{-\frac{2\alpha+2x}{\alpha}}, r<r_{\rm ext} \\
{\nu}^{-\frac{2\alpha+2x-n}{\alpha}}, r>r_{\rm ext}
\end{array} \right.
\end{equation}
in the comoving frame, and 
\begin{equation}\label{Eq17}
Y\propto \left\{
\begin{array}{rl}
{\nu^{\rm obs}}^{\frac{2\alpha+2x}{x-\alpha}}, r<r_{\rm ext} \\
{\nu^{\rm obs}}^{\frac{2\alpha+2x-n}{x-\alpha}}, r>r_{\rm ext}
\end{array} \right.
\end{equation}
in the observers' frame. For the correlation between the synchrotron peak frequency and the radiative luminosity, we have
\begin{equation}\label{Eq18}
\begin{split}
L_{\rm rad}\propto\left\{
\begin{array}{rl}
&\nu, \it{Y}\ll1 \\
&{\nu}^{-\frac{2x+\alpha}{\alpha}}, \it{Y}\gg1, r<r_{\rm ext} \\
&{\nu}^{-\frac{2x+\alpha-n}{\alpha}}, \it{Y}\gg1, r>r_{\rm ext}
\end{array} \right.
\end{split}
\end{equation}
in the comoving frame, and 
\begin{equation}\label{Eq19}
\begin{split}
L_{\rm rad}^{\rm obs}\propto\left\{
\begin{array}{rl}
&{\nu^{\rm obs}}^{\frac{4x-\alpha}{x-\alpha}}, \it{Y}\ll1 \\
&{\nu^{\rm obs}}^{\frac{6x+\alpha}{x-\alpha}}, \it{Y}\gg1, r<r_{\rm ext} \\
&{\nu^{\rm obs}}^{\frac{6x+\alpha-n}{x-\alpha}}, \it{Y}\gg1, r>r_{\rm ext}
\end{array} \right.
\end{split}
\end{equation}
in the observers' frame.

For BL Lacs, the Compton dominance $Y$ can be evaluated as 
\begin{equation}
Y\simeq \frac{U_{\rm syn}}{U_B},
\end{equation}
where $U_{\rm syn}=L_{\rm rad}^{\rm syn}/(4\pi R^2c)$ is the energy density of synchrotron photons, $L_{\rm rad}^{\rm syn}$ represents the radiative synchrotron luminosity 
\begin{equation}\label{L_radB}
L_{\rm rad}^{\rm syn}=V_{\rm b}mc^2\int_{\gamma_{\rm min}}^{\gamma_{\rm max}} f_{\rm syn}(\gamma)\gamma \dot{Q}(\gamma)d\gamma,
\end{equation}
where $f_{\rm syn}(\gamma)=\rm min\{\frac{\it{t}_{\rm dyn}}{\it{t}_{\rm syn}}, 1\}$ is the cooling efficiency of synchrotron emission. Then $Y$ can be written as
\begin{equation}\label{Y_B}
Y\propto L_{\rm inj}(\gamma>\gamma_{\rm c})+L_{\rm inj}(\gamma<\gamma_{\rm c})r^{-\alpha}.
\end{equation}
Firstly, we study the correlations of the blazar sequence for $\nu=\nu_{\rm c}$. If $Y$'s the first term on the right is dominant, the correlation between $\nu$ and $Y$ depends on if there is a correlation between $\nu$ and $L_{\rm inj}$. If the first term on the right is not dominant, we derive the correlation between $\nu$ and $Y$,
\begin{equation}\label{BY<1}
\begin{split}
Y\propto \left\{
\begin{array}{rl}
&{\nu}^{-1}, Y\ll1 \\
&{\nu}^{-\frac{1}{3}}, \it{Y}\gg1
\end{array} \right.
\end{split}
\end{equation}
in the comoving frame, and  
\begin{equation}\label{BY>1}
\begin{split}
Y\propto \left\{
\begin{array}{rl}
&{\nu^{\rm obs}}^{-\frac{\alpha}{\alpha+x}}, Y\ll1 \\
&{\nu^{\rm obs}}^{-\frac{\alpha}{3\alpha+x}}, \it{Y}\gg1
\end{array} \right.
\end{split}
\end{equation}
in the observers' frame.
Similarly, we derive the correlation between the synchrotron peak frequency and the radiative luminosity, i.e.,
\begin{equation}\label{Eq28}
\begin{split}
L_{\rm rad}\propto\left\{
\begin{array}{rl}
&{\nu}^{-1}, \it{Y}\ll1 \\
&{\nu}^{-2/3}, \it{Y}\gg1
\end{array} \right.
\end{split}
\end{equation}
in the comoving frame, and 
\begin{equation}\label{Eq29}
\begin{split}
L_{\rm rad}^{\rm obs}\propto\left\{
\begin{array}{rl}
&{\nu^{\rm obs}}^{\frac{4x-\alpha}{x+\alpha}}, \it{Y}\ll1 \\
&{\nu^{\rm obs}}^{\frac{4x-2\alpha}{x+3\alpha}}, \it{Y}\gg1
\end{array} \right.
\end{split}
\end{equation}
in the observers' frame.

Similar to previous discussion for FSRQs, we also study these two correlations for BL Lacs in the case of $\nu<\nu_{\rm c}$, i.e., $\nu\propto B^3r^2\propto r^{-\alpha}$. For the correlation between the synchrotron peak frequency and the Compton dominance (Eq.~\ref{Y_B}), we have
\begin{equation}\label{Eq30}
Y\propto {\nu}
\end{equation}
in the comoving frame, and  
\begin{equation}\label{Eq31}
Y\propto {\nu^{\rm obs}}^{\frac{\alpha}{\alpha-x}}
\end{equation}
in the observers' frame.
For the correlation between the synchrotron peak frequency and the radiative luminosity, we have
\begin{equation}\label{BLC}
\begin{split}
L_{\rm rad}\propto\left\{
\begin{array}{rl}
&\nu, \it{Y}\ll1 \\
&\nu^{2}, \it{Y}\gg1
\end{array} \right.
\end{split}
\end{equation}
in the comoving frame, and
\begin{equation}\label{BLO}
\begin{split}
L_{\rm rad}^{\rm obs}\propto\left\{
\begin{array}{rl}
&{\nu^{\rm obs}}^{\frac{4x-\alpha}{x-\alpha}}, \it{Y}\ll1 \\
&{\nu^{\rm obs}}^{\frac{4x-2\alpha}{x-\alpha}}, \it{Y}\gg1
\end{array} \right.
\end{split}
\end{equation}
in the observers' frame. 

\begin{table*}
\caption{Deduced equations of the correlations in the blazar sequence.}\label{table0}
\centering
\resizebox{2.2\columnwidth}{!}{
\renewcommand{\arraystretch}{1.4}
\Huge
\begin{tabular}{cccccc}
\hline\hline							
$\nu=\nu_{\rm c}$											
	&	in the observers' frame	&	in the comoving frame	&		&	in the observers' frame	&	in the comoving frame	\\
\hline											
FSRQs, $Y\ll1$	&		&		&	BL Lacs, $Y\ll1$	&		&		\\
\hline
	&	Eq.~(\ref{Eq12}): $ Y\propto \left\{
\begin{array}{rl}
{\nu^{\rm obs}}^2, r<r_{\rm ext} \\
{\nu^{\rm obs}}^{\frac{2\alpha+2x-n}{\alpha+x}}, r>r_{\rm ext}
\end{array} \right. $	&	Eq.~(\ref{FY<1}): $ Y\propto \left\{
\begin{array}{rl}
{\nu}^{\frac{2\alpha+2x}{\alpha}}, r<r_{\rm ext} \\
{\nu}^{\frac{2\alpha+2x-n}{\alpha}}, r>r_{\rm ext}
\end{array} \right. $	&		&	Eq.~(\ref{BY>1}):  $Y\propto {\nu^{\rm obs}}^{-\frac{\alpha}{\alpha+x}}$	&	Eq.~(\ref{BY<1}): $Y\propto {\nu}^{-1}$	\\
	&	Eq.~(\ref{FLO}): $ L_{\rm rad}^{\rm obs}\propto {\nu^{\rm obs}}^{\frac{4x-\alpha}{x+\alpha}}$	&	Eq.~(\ref{FLC}): $L_{\rm rad}\propto {\nu}^{-1}$	&		&	Eq.~(\ref{Eq29}): $L_{\rm rad}^{\rm obs}\propto {\nu^{\rm obs}}^{\frac{4x-\alpha}{x+\alpha}}$	&	Eq.~(\ref{Eq28}): $L_{\rm rad}\propto {\nu}^{-1}$	\\
\hline											
FSRQs, $Y\gg1$	&		&		&	BL Lacs, $Y\gg1$	&		&		\\
\hline											
	&	Eq.~(\ref{Eq12}): $Y\propto \left\{
\begin{array}{rl}
{\nu^{\rm obs}}^{-2/3}, r<r_{\rm ext} \\
{\nu^{\rm obs}}^{\frac{2\alpha+2x-n}{2n-3\alpha-3x}}, r>r_{\rm ext}
\end{array} \right. $	&	Eq.~(\ref{FY<1}): $Y\propto \left\{
\begin{array}{rl}
{\nu}^{\frac{2\alpha+2x}{-3\alpha-4x}}, r<r_{\rm ext} \\
{\nu}^{\frac{2\alpha+2x-n}{2n-3\alpha-4x}}, r>r_{\rm ext}
\end{array} \right.$	&		&	Eq.~(\ref{BY>1}): $Y\propto {\nu^{\rm obs}}^{-\frac{\alpha}{3\alpha+x}}$	&	Eq.~(\ref{BY<1}): $Y\propto {\nu}^{-\frac{1}{3}}$	\\
	&	Eq.~(\ref{FLO}): $L_{\rm rad}^{\rm obs}\propto\left\{
\begin{array}{rl}
&{\nu^{\rm obs}}^{\frac{6x+\alpha}{-3\alpha-3x}}, r<r_{\rm ext} \\
&{\nu^{\rm obs}}^{\frac{6x+\alpha-n}{2n-3\alpha-3x}}, r>r_{\rm ext}
\end{array} \right.$	&	Eq.~(\ref{FLC}): $L_{\rm rad}\propto\left\{
\begin{array}{rl}
&{\nu}^{\frac{2x+\alpha}{-3\alpha-4x}}, r<r_{\rm ext} \\
&{\nu}^{\frac{2x+\alpha-n}{2n-3\alpha-4x}}, r>r_{\rm ext}
\end{array} \right.$	&		&	Eq.~(\ref{Eq29}): $L_{\rm rad}^{\rm obs}\propto {\nu^{\rm obs}}^{\frac{4x-2\alpha}{x+3\alpha}}$	&	Eq.~(\ref{Eq28}): $L_{\rm rad}\propto {\nu}^{-2/3}$	\\
\hline\hline											
$\nu<\nu_{\rm c}$											
	&	&		&		&	&		\\
\hline											
FSRQs	&		&		&	BL Lacs	&		&		\\
\hline
	&	Eq.~(\ref{Eq17}): $Y\propto \left\{
\begin{array}{rl}
{\nu^{\rm obs}}^{\frac{2\alpha+2x}{x-\alpha}}, r<r_{\rm ext} \\
{\nu^{\rm obs}}^{\frac{2\alpha+2x-n}{x-\alpha}}, r>r_{\rm ext}
\end{array} \right.$	&	Eq.~(\ref{Eq16}): $Y\propto \left\{
\begin{array}{rl}
{\nu}^{-\frac{2\alpha+2x}{\alpha}}, r<r_{\rm ext} \\
{\nu}^{-\frac{2\alpha+2x-n}{\alpha}}, r>r_{\rm ext}
\end{array} \right.$	&		&	Eq.~(\ref{Eq31}): $Y\propto {\nu^{\rm obs}}^{\frac{\alpha}{\alpha-x}}$	&	Eq.~(\ref{Eq30}): $Y\propto {\nu}$	\\
	&	Eq.~(\ref{Eq19}): $L_{\rm rad}^{\rm obs}\propto\left\{
\begin{array}{rl}
&{\nu^{\rm obs}}^{\frac{4x-\alpha}{x-\alpha}, \it{Y}\ll1} \\
&{\nu^{\rm obs}}^{\frac{6x+\alpha}{x-\alpha}, \it{Y}\gg1}, r<r_{\rm ext} \\
&{\nu^{\rm obs}}^{\frac{6x+\alpha-n}{x-\alpha}, \it{Y}\gg1}, r>r_{\rm ext}
\end{array} \right.$	&	Eq.~(\ref{Eq18}): $L_{\rm rad}\propto\left\{
\begin{array}{rl}
&\nu, \it{Y}\ll1 \\
&{\nu}^{-\frac{2x+\alpha}{\alpha}}, \it{Y}\gg1, r<r_{\rm ext} \\
&{\nu}^{-\frac{2x+\alpha-n}{\alpha}}, \it{Y}\gg1, r>r_{\rm ext}
\end{array} \right.$	&		&	Eq.~(\ref{BLO}): $L_{\rm rad}^{\rm obs}\propto\left\{
\begin{array}{rl}
&{\nu^{\rm obs}}^{\frac{4x-\alpha}{x-\alpha}}, \it{Y}\ll1 \\
&{\nu^{\rm obs}}^{\frac{4x-2\alpha}{x-\alpha}}, \it{Y}\gg1
\end{array} \right.$	&	Eq.~(\ref{BLC}): $L_{\rm rad}\propto\left\{
\begin{array}{rl}
&\nu, \it{Y}\ll1 \\
&\nu^{2}, \it{Y}\gg1
\end{array} \right.$	\\
\hline\hline																							
\end{tabular}}
\end{table*}

As deduced above, we obtain equations of the blazar sequence for FSRQs and for BL Lacs in the comoving and observers' frames, respectively. All equations are summarized in Table~\ref{table0}. By taking the logarithm of the parameters in the correlations, we can get the corresponding slopes, which can be compared with the statistical linear regression results. 

The simple scaling model proposed in this subsection is quite similar to the conventional one-zone model, where all the jet emission is represented by one single dissipation region. Compared to the conventional one-zone model, the advantage of this simple scaling model is that power-law relationships have been established among several physical parameters based on reasonable assumptions. However, normalizations of power-law relationships for each blazar remains unknown and varies. This will greatly increase the dispersion of the correlation intercept, even if the correlation slope predicted by the model is the same. Moreover, the simple scaling model, compared to the conventional one-zone model, takes many shortcuts. For example, to establish the power-law correlations among multiple physical parameters, a crude assumption is introduced, i.e., the electron injected luminosities of all blazars do not have a large dispersion. However, as suggested by \cite{2008MNRAS.387.1669G}, the black hole mass and the accretion rate would be crucial for interpreting the blazar sequence, since the jet power is always linked with the accretion rate. Therefore, such an assumption is likely to generate potential bias. Also, blazars have a complex environment with various external photon fields, such as the BLR, the DT, etc, each with its own unique influence distance. Gathering information on each blazar's external photon field is tough, so we make an simplification that there is just one main external photon field. When using the conventional one-zone model to interpret blazars' SEDs, the evolution of the relativistic electron energy distribution is a factor that cannot be ignored. However, in our attempt to understand the blazar sequence using the simple scaling model, we do not specify the particular shape for the electron energy distribution, nor do we study its evolution. This simplification is feasible because the study of the blazar sequence only requires information about the synchrotron peak frequency and integrated luminosity, both of which can be obtained through analytical calculations.

\subsection{Theoretical Implications}\label{TD}
In this subsection, we apply the simple scaling model described in Sect.~\ref{SDM} to interpret the statistical correlation results of the phenomenological and intrinsic blazar sequence derived in Sect.~\ref{revisit}.

First of all, let us investigate if statistical results can be interpreted under the conditions that $\nu=\nu_{\rm c}$, i.e., cooling is important. As mentioned before, values of $\delta_{\rm D}$ given by radio observation may have a great discrepancy with the actual values of $\delta_{\rm D}$. If one consider that radio $\delta_{\rm D}$ is significantly influential and unreliable, or if we believe that the distribution of actual $\delta_{\rm D}$ is irregular, i.e., $x=0$, then it can be seen from Table~\ref{table0} that the synchrotron peak frequency is both anti-correlated with the Compton dominance and the radiative luminosity in the observers' and comoving frames. This is consistent with the original physical interpretation proposed by \cite{1998MNRAS.301..451G}. If here we trust the radio $\delta_{\rm D}$ and believe in the correlation results of the intrinsic blazar sequence, then the physical explanation of cooling might faces difficulties. From the statistical results in Table~\ref{table1}, it can be seen that no correlation is found between the Compton dominance and the synchrotron peak frequency for all (sub-)samples of FSRQs, whether in the observers' frame or the comoving frame. It indicates that indexes of the deduced Eqs.~(\ref{FY<1}, \ref{Eq12}) are zero, i.e., $r>r_{\rm ext}$ and $2\alpha+2x-n=0$. Consequently, the indexes of Eq.~(\ref{FLC}) between $L_{\rm rad}$ and $\nu$ in the comoving frame become -1. In addition, Eq.~(\ref{Eq28}) suggests that $L_{\rm rad}$ is negatively correlated with $\nu$ for BL Lacs in the comoving frame (even when considering the actual nonlinear correlation). However, moderate and strong positive correlations between $L_{\rm rad}$ and $\nu$ are found for FSRQs and BL Lacs, respectively, implying a different physical origin of these correlations. In the following, we will discuss if deduced equations under the condition that $\nu<\nu_{\rm c}$ have the potential to account for the phenomenological and intrinsic blazar sequence simultaneously.

\begin{figure}
\centering
\includegraphics[width=9cm,height=6cm]{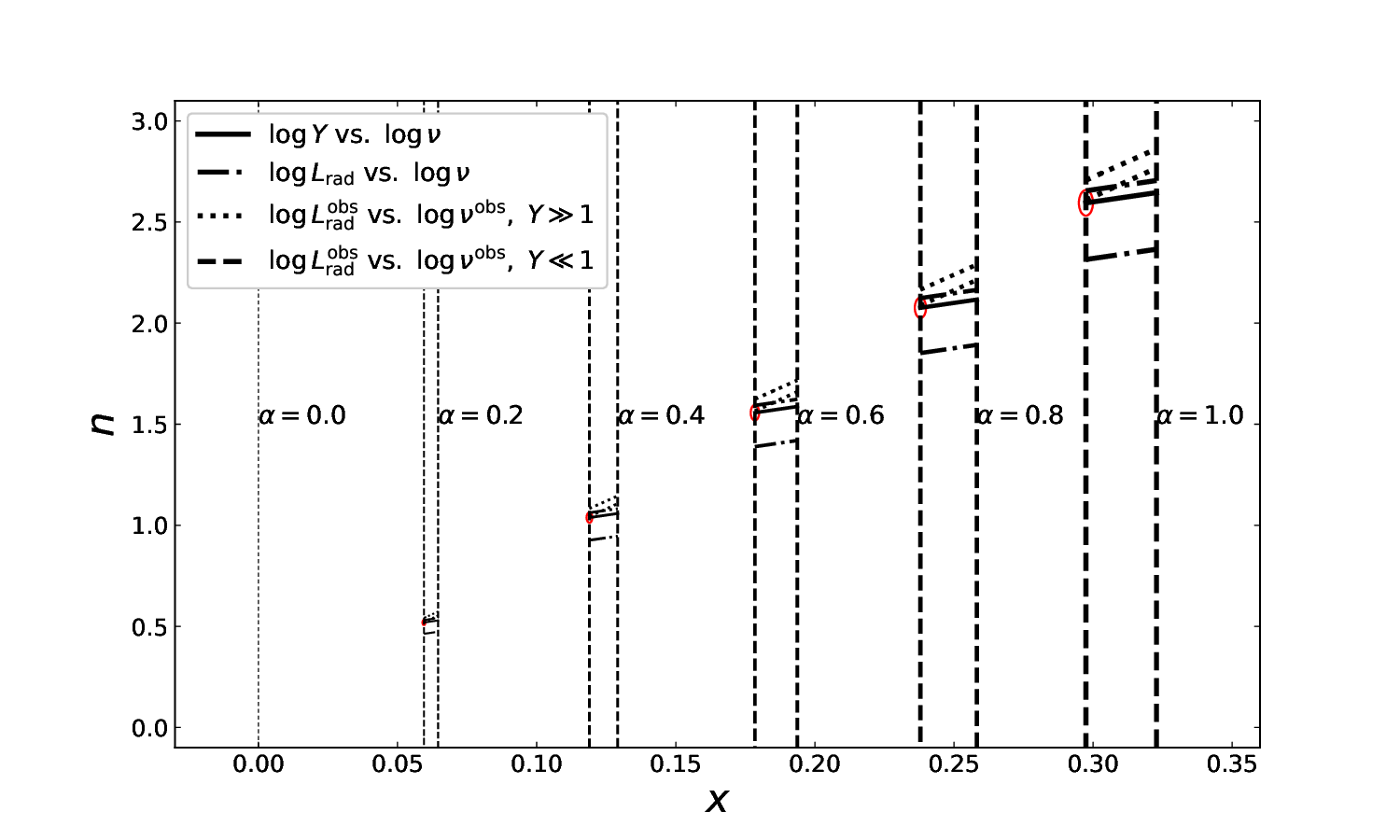}
\caption{Parameter spaces for FSRQs with different value of $\alpha$ when $r>r_{\rm ext}$. Solid curves represent that the indexes of Eqs.~(\ref{Eq16}, \ref{Eq17}) are equal to zero, as statistical analysis find that $Y$ and $\nu$ (or $\nu^{\rm obs}$) are not correlated. Dashed, dotted, and dash-dotted curves are obtained by setting the indexes of Eqs.~(\ref{Eq18}, \ref{Eq19}) equal to slopes obtained from statistical analysis, respectively. Since errors of slopes are also taken into account, the intersection area between two curves of each type represents the effective parameter space. After considering the intersection of all line types, we find the corresponding combinations of $n$, $x$, and $\alpha$, which are marked by red circles.
}
\label{Fnxa}
\end{figure}

For FSRQs, if we consider $r<r_{\rm ext}$, Eqs.~(\ref{Eq16}, \ref{Eq17}) suggest that $x$ should be equal to $-\alpha$, since no correlation is found between the Compton dominance and the synchrotron peak frequency both in the observers' frame and the comoving frame. Consequently, the index of Eq.~(\ref{Eq18}) becomes unity (including the case $Y\ll1$), which is consistent with the statistical results. However, the index of Eq.~(\ref{Eq19}) will be 5/2, which is inconsistent with the statistical results that found $s=-0.35$ when $Y\leq1$ and $s=-0.18$ when $Y>1$. Therefore, we need to further discuss the case of $r>r_{\rm ext}$. In this case, since statistical studies find that the Compton dominance and the synchrotron peak frequency are not correlated, Eqs.~(\ref{Eq16}, \ref{Eq17}) suggest that $2\alpha + 2x - n = 0$. Moreover, by making the indexes of Eqs.~(\ref{Eq18}, \ref{Eq19}) equal to the slopes obtained in the statistical studies (including errors), we can get combinations of $x$, $n$, and $\alpha$ that satisfies all conditions, which are shown in Figure~\ref{Fnxa}. It can be seen that the size of effective intersection area is proportional to the value of $\alpha$. It implies that the effective parameter space can only be obtained when $x>0$, i.e., the dissipation region has to be located in the accelerating jet. We emphasize that any combinations of $x$, $n$, and $\alpha$ that satisfies $2\alpha + 2x - n = 0$ are possible. In Figure~\ref{Fnxa}, we present five sets of parameter combinations, which are $\alpha=1, n\simeq2.6, x\simeq0.3$; $\alpha=0.8, n\simeq2.1, x\simeq0.24$; $\alpha=0.6, n\simeq1.56, x\simeq0.18$; $\alpha=0.4, n\simeq1.04, x\simeq0.12$; and $\alpha=0.2, n\simeq0.52, x\simeq0.06$.
Please note that our discussion here is based on the assumption of a single external photon field, which is a simplification. In the actual AGNs' environment, it is generally believed that there are two external photon fields, i.e., the BLR and the DT, both of which could have significant implications for jet emission. For the BLR, $n=3$ has been suggested as a model assumption by \cite{2009ApJ...704...38S}. For the DT, $n=4$ has been found by \cite{2012ApJ...754..114H} for a specific observation of an FSRQ 3C 279. While it may not be the case that $n=3$ or $n=4$ works for every AGN, our results clearly suggest that the majority of AGNs cannot have an $n$ greater than 3. Otherwise, $\alpha$ would be larger than 1, implying that the jet profile becomes hyperbolic. This is evidently in contradiction with the conical or parabolic structure found by radio observations \citep[see][for a review]{2019ARA&A..57..467B}. On the other hand, the covering factors, which represent the fractions of the disk luminosity reprocessed into the BLR and DT radiation, evidently vary among different AGNs. For instance, values such as 0.1, 0.2, and 0.5 have been suggested \citep{2007A&A...468..979M, 2009MNRAS.397..985G, 2010ApJ...724L..59H, 2011ApJ...733..108H}. This degree of variability introduces complexity into the model interpretation attempted here, and may even preclude the possibility of reaching definitive conclusions. Therefore, the assumption of a single external photon field in this work is a simplifying approximation. In general, our model suggests that the phenomenological and intrinsic blazar sequence of FSRQs can be explained only when the condition $2\alpha + 2x - n = 0$ is satisfied, implying that the dissipation region is in an accelerating jet located beyond the external photon field in the slow-cooling regime.

For BL Lacs, the physical interpretation of phenomenological and intrinsic blazar sequence appears to be complicated. For BL Lacs with $Y>1$, the derived correlation results and slopes (if correlations exist) are similar to those of FSRQs. On the other hand, we find that BL Lacs with $Y>1$ in different samples are dominated by LSPs (61/118 = 52 \% for historical sample, 14/17 = 82 \% for quasi-simultaneous sample, and 23/51 = 45 \% for $\delta_{\rm D}$-corrected sample), which may suggest that their high-energy components are mainly from the EC emission rather than the SSC emission \citep{2007Ap&SS.309...95B, 2013ApJ...768...54B, 2021RAA....21..305W, 2023MNRAS.521.6210D}. 
Blazars are classified as FSRQs and BL Lacs based on whether broad emission lines are detected. However, there are many blazars with comparable jet and broad emission lines intensities that are classified into either subclass depending on the jet activity during observation. For instance, if the jet emission is in a low state during observation, the blazar will be observed with broad emission lines and classified as FSRQs. On the other hand, if the jet emission is in a high state, the emission lines will be masked, then the blazar will be classified as BL Lacs. Such blazars are known as ``changing-look'' blazars \citep{2003MNRAS.342..422M, 2005A&A...442..185B}. Additionally, some blazars with broad emission lines are classified as BL Lacs because their broad emission lines are outshone by the jet emission. These blazars are suggested as ``masquerading'' BL Lacs \citep{2013MNRAS.431.1914G}. In our sample, eighteen blazars (including 4FGL J0238.6+1637, 4FGL J0334.2-4008, 4FGL J0407.5+0741, 4FGL J0428.6-3756, 4FGL J0438.9-4521, 4FGL J0516.7-6207, 4FGL J0538.8-4405, 4FGL J0629.3-1959, 4FGL J0710.9+4733, 4FGL J0831.8+0429, J1001.1+2911, 4FGL J1058.4+0133, 4FGL J1147.0-3812, 4FGL J1751.5+0938, 4FGL J1800.6+7828, 4FGL J1954.6-1122, 4FGL J2134.2-0154, and 4FGL J2152.5+1737) are suggested as changing-look blazars \citep{2022ApJ...936..146X}, which can be seen as the direct evidence of presence of external photon fields. Therefore, we suggest that the phenomenological and intrinsic blazar sequence of BL Lacs with $Y>1$ can also be explained by the dissipation region in accelerating jet (located beyond the external photon field) emitting in the slow-cooling regime, as in FSRQs. For BL Lacs with $Y\le1$, we do not find enough BL Lacs with measured $\delta_{\rm D}$ in the literature, so we only study the phenomenological blazar sequence. For the phenomenological correlations, no correlation is found between $ \log (L_{{\rm rad}}^{{\rm obs}})~ {\rm vs.}~ \log (\nu^{{\rm {obs}}})$, indicating the indexes of Eq.~(\ref{Eq29}, \ref{BLO}) are zeros, i.e., $4x-\alpha=0$. Since $0<\alpha \leq 1$, it indicates that the dissipation region is in an accelerating jet with $0<x\leq0.25$ (no matter $\nu=\nu_{\rm c}$ or $\nu<\nu_{\rm c}$). If defaulting $x=0.25$, the obtained strong negative correlation between $\log (Y)$ and $\log (\nu^{{\rm {obs}}})$ can be explained in the case of $\nu=\nu_{\rm c}$. Such a cooling scenario is quite possible for BL Lacs with $Y\le1$ since HSPs are dominant in the historical (472/726=65\%) and quasi-simultaneous (5/11=45\%) samples. On the other hand, it is necessary to note that the model predicted slope ($-0.8$) for the correlation between $\log (Y)$ and $\log (\nu^{{\rm {obs}}})$ is lower than that derived by the statistical analysis ($\sim -0.2$). We suppose that this might be due to the BL Lacs in two samples having mixed cooling regimes, since the model predicted correlations under two cooling scenarios are opposite as shown in Table~\ref{table0}. Here we emphasize that the above discussion assumes $x=0.25$ merely for the convenience of showing that the statistical correlations could be reproduced based on some specific conditions. In fact, any combination of $x$ and $\alpha$ that satisfies $4x-\alpha=0$ is plausible. Furthermore, the attemp of introducing mixed cooling scenarios arises from the single cooling scenario's inadequacy to account for the slope derived from statistical analysis.

\subsection{Reproducing the blazar sequence}\label{RBS}
In previous, our theoretical analysis of the blazar sequence focuses on the power-law relation between various parameters (summarized in Table~\ref{table0}), as the obtained indexes can be compared with slopes given by correlation studies. However, specific values of normalization in power-law relations are ignored, which will inevitably increase the dispersion of the correlation. In addition, we do not provide specific values for each physical parameter, so one may worry that special and unreasonable physical parameters will be introduced. On the other hand, as shown in Sect.~\ref{SDM}, the two correlations of the blazar sequence under some conditions are nonlinear, which will have a certain impact on the slope and dispersion of the correlation. In this subsection, we would like to apply the simple scaling model to reproduce the blazar sequence shown in Figures~\ref{correlations_Y}--\ref{correlations_corrected sample}, and provide the distribution of important physical parameters. In the following, we default that the jet has a conical structure (i.e., $\alpha=1$) as the benchmark case, as found in observations \citep{2007ApJ...668L..27K, 2011A&A...532A..38S} and assumed in theoretical models \citep{2006MNRAS.367.1083K, 2012MNRAS.423..756P}.

\begin{figure*}
\centering
\subfloat{
\includegraphics[width=0.7\columnwidth]{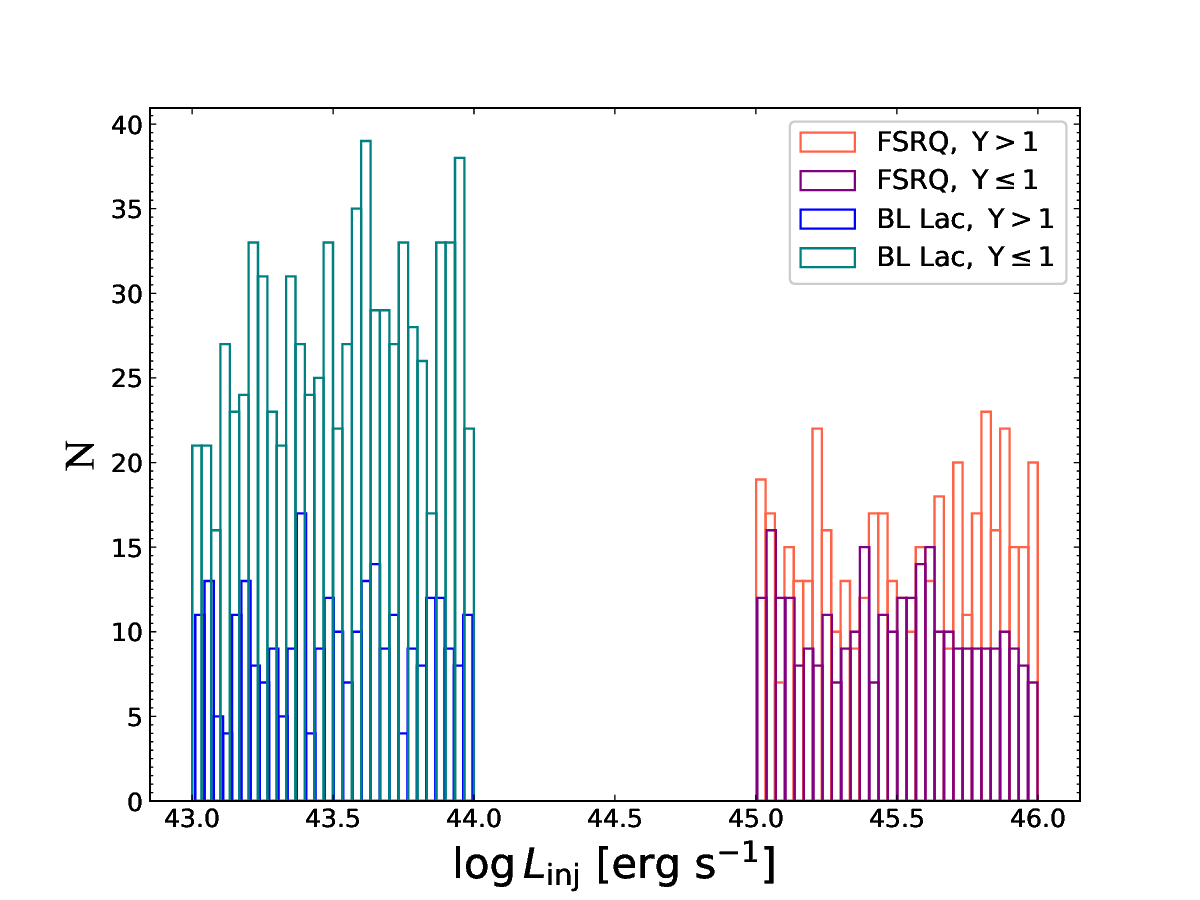}
}\hspace{-9mm}
\vspace{-1mm}
\quad
\subfloat{
\includegraphics[width=0.7\columnwidth]{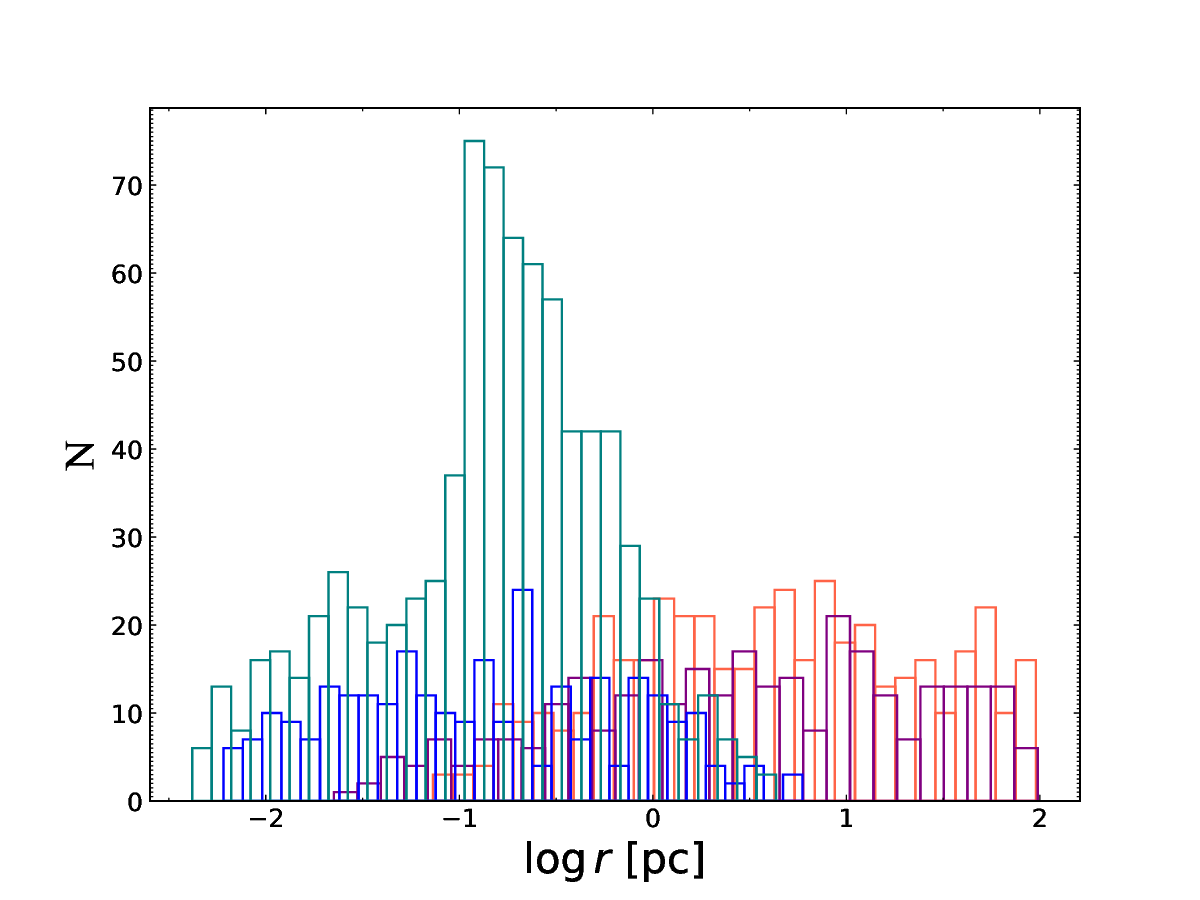}
}\hspace{-9mm}
\vspace{-1mm}
\quad
\subfloat{
\includegraphics[width=0.7\columnwidth]{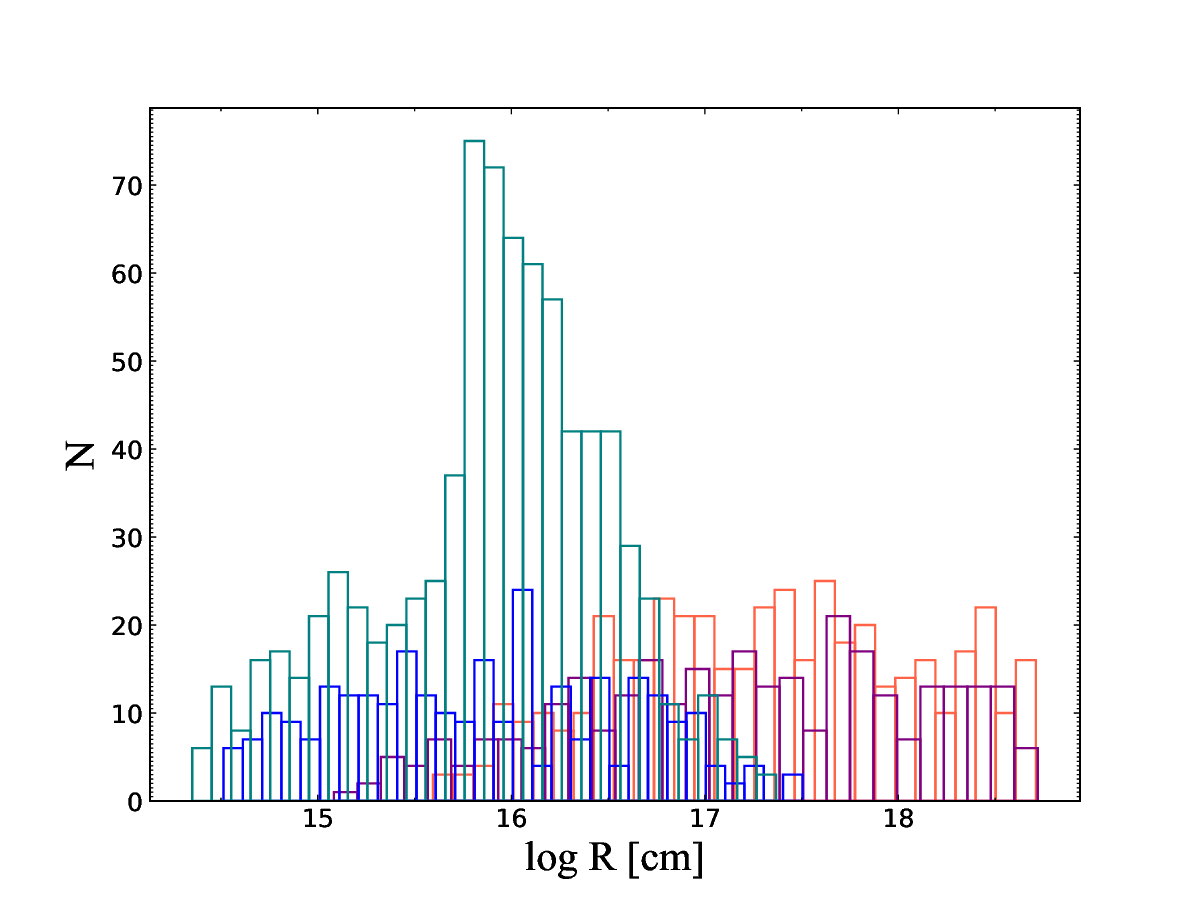}
}\hspace{-9mm}
\vspace{-1mm}
\quad
\subfloat{
\includegraphics[width=0.7\columnwidth]{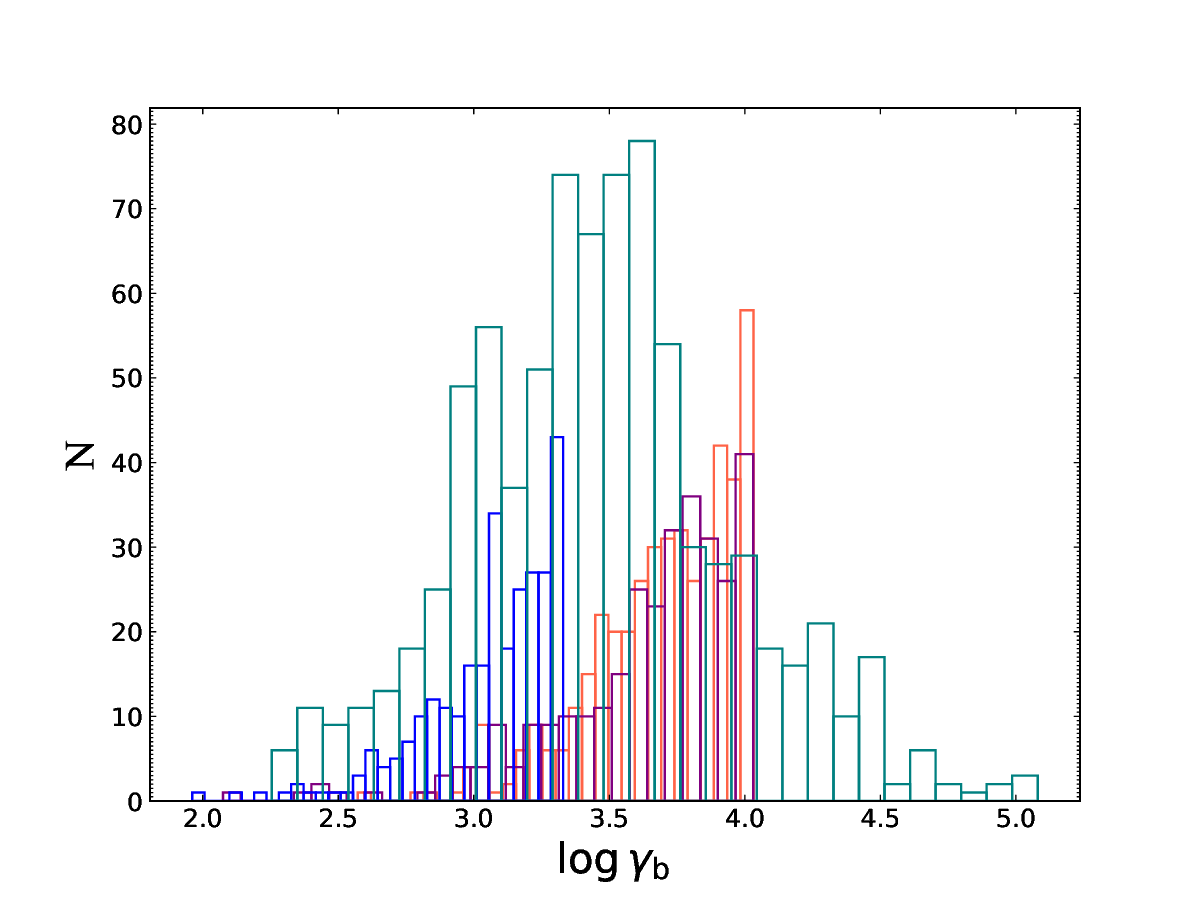}
}\hspace{-9mm}
\vspace{-1mm}
\quad
\subfloat{
\includegraphics[width=0.7\columnwidth]{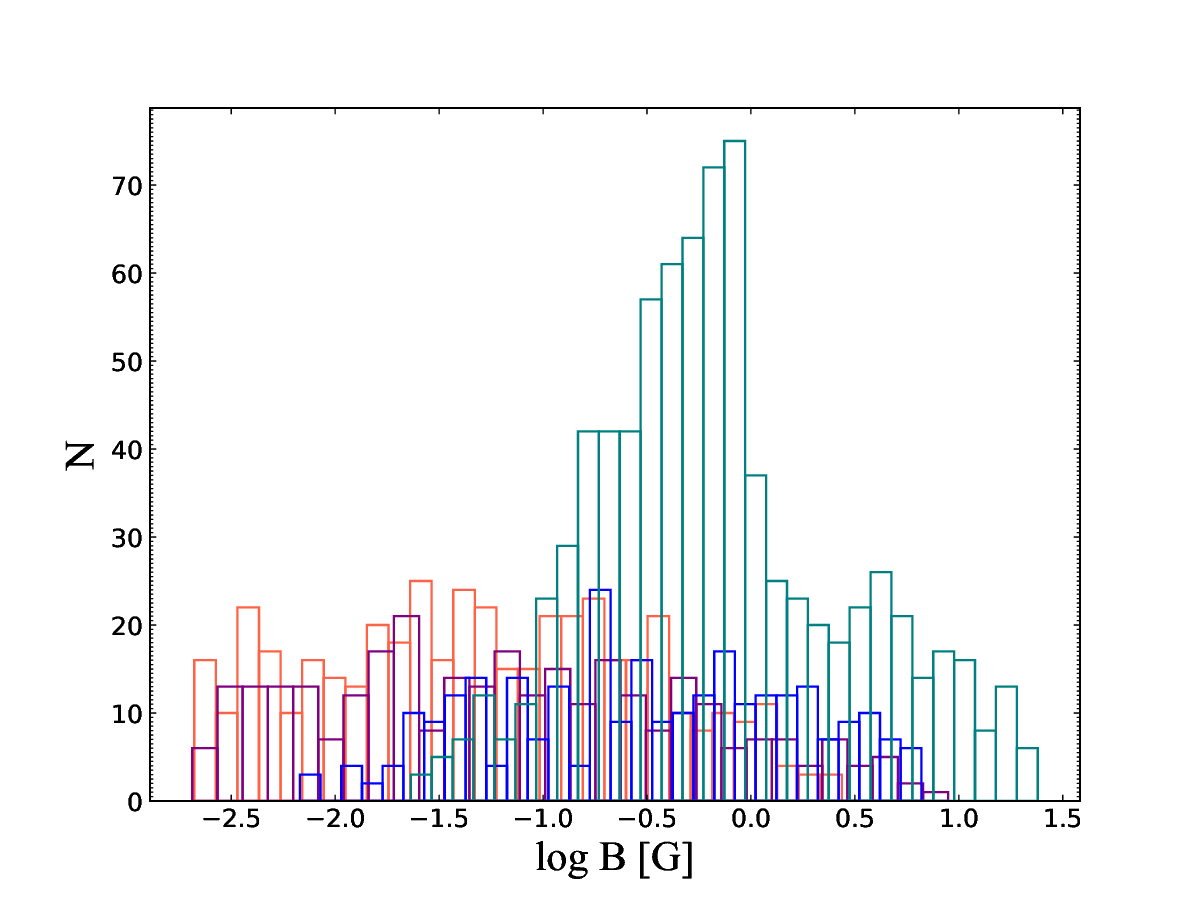}
}\hspace{-9mm}
\vspace{-1mm}
\quad
\subfloat{
\includegraphics[width=0.7\columnwidth]{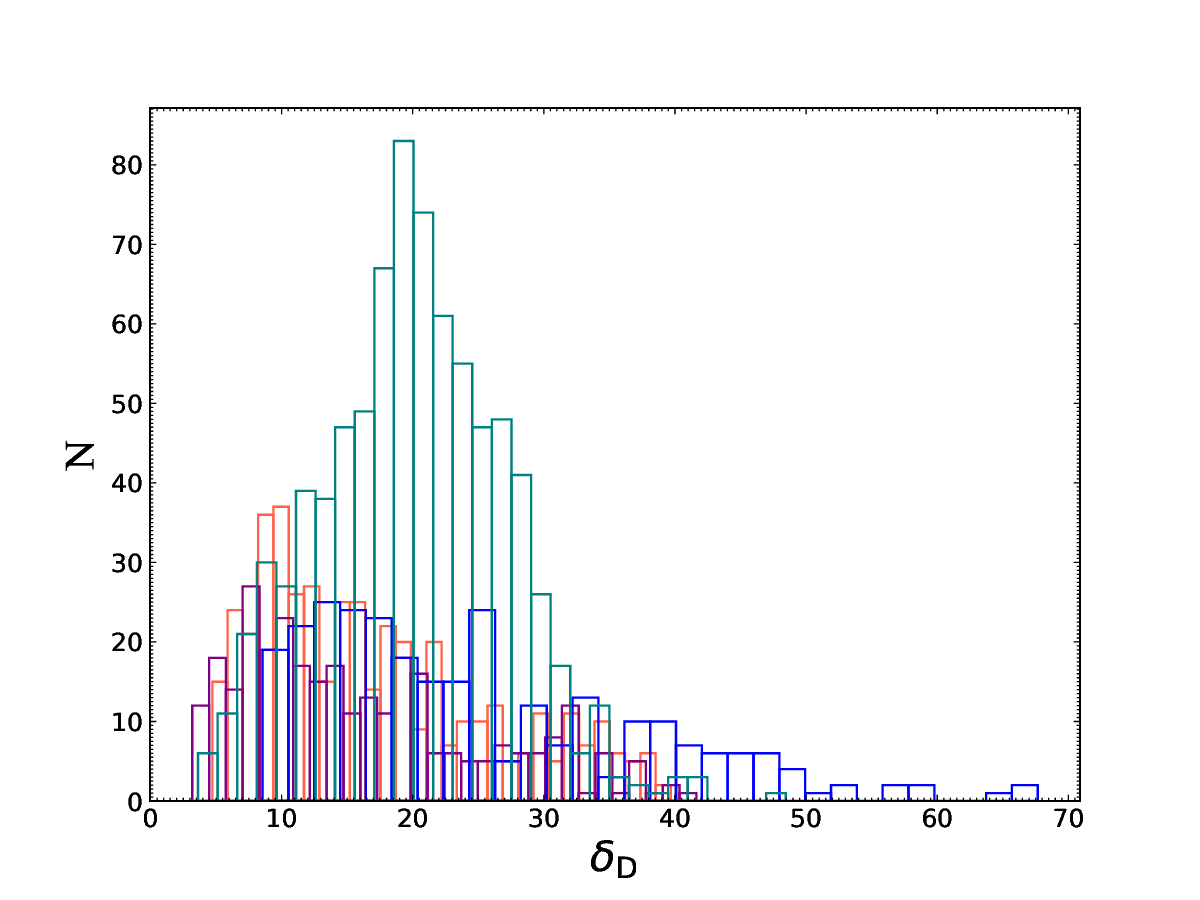}
}\hspace{-9mm}
\vspace{-1mm}
\quad
\subfloat{
\includegraphics[width=0.7\columnwidth]{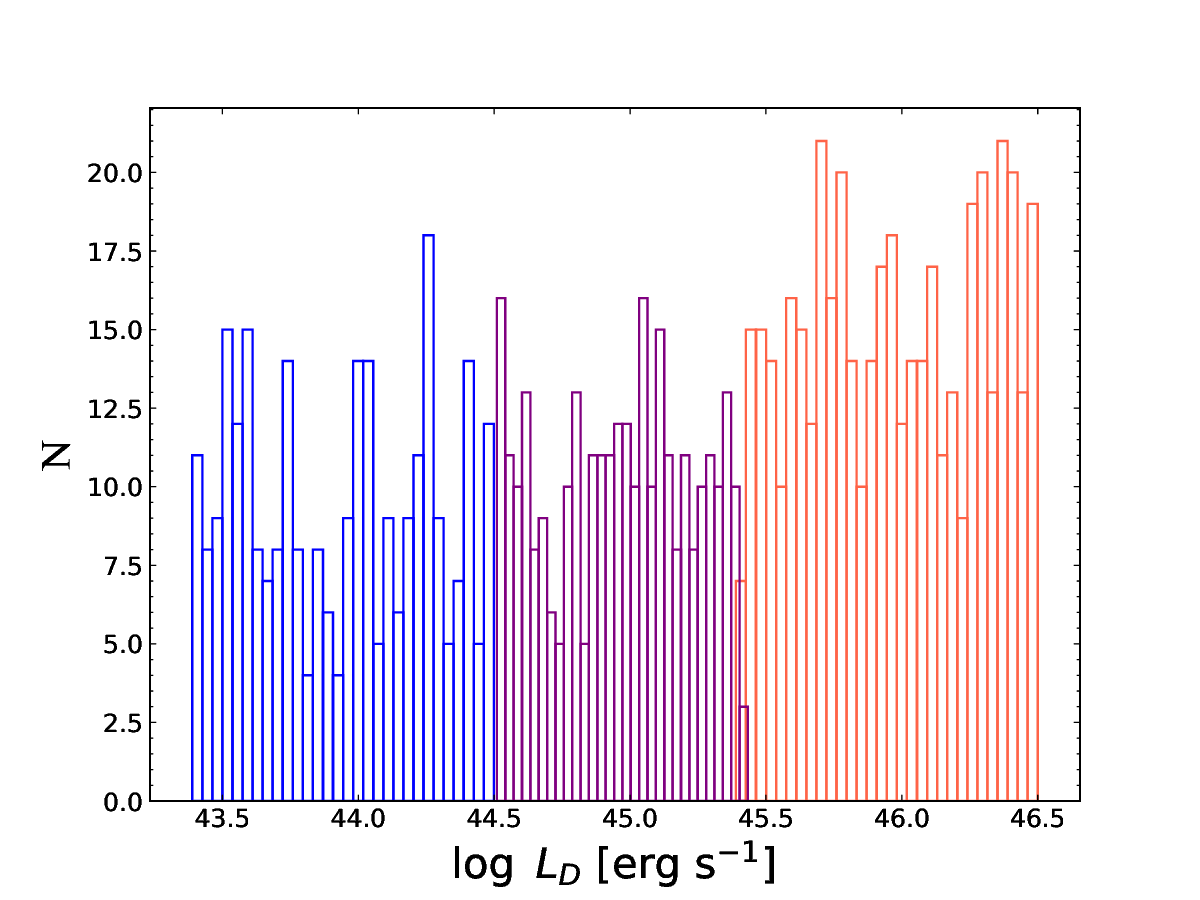}
}
\caption{Distributions of the electron injection luminosity $ \log L_{\rm inj}$, the distance between SMBH
and the dissipation region $ \log r$, the radius of the dissipation region $ \log R$, the electron Lorentz factor corresponding to the synchrotron peak frequency $\log \gamma_{\rm b}$, the magnetic field in the dissipation region $ \log B$, the Doppler factor of the dissipation region $\log \delta_{\rm D}$, and the disk luminosity $\log L_{\rm D}$. Mean values for FSRQs are, $\bar{L}_{\rm inj}=3.2\times10^{45}\rm~erg~s^{-1}$, $\bar{r}=3.5~\rm pc$, $ \bar{R}=1.9\times10^{17}\rm~cm$, $\bar{\gamma}_{\rm b}=4.8\times10^3$, $\bar{B}=0.3\rm~G$, $\bar{\delta}_{\rm D}=16.8$, $\bar{L}_{\rm D}=3.6\times10^{45}\rm~erg~s^{-1}$, respectively. Mean values for BL Lacs are, $\bar{L}_{\rm inj}=3.3\times10^{43}\rm~erg~s^{-1}$, $\bar{r}=0.14\rm~pc$, $ \bar{R}=7.8\times10^{15}\rm~cm$, $\bar{\gamma}_{\rm b}=2.3\times10^3$, $\bar{B}=1.7\rm~G$, $\bar{\delta}_{\rm D}=21.2$, $\bar{L}_{\rm D}=8.7\times10^{43}\rm~erg~s^{-1}$, respectively. The meanings of histograms with different colors are explained in the inset legends.\label{para}}
\end{figure*}

Applying the simple scaling model, 1800 blazars, including 300 FSRQs with $Y\leq1$, 500 FSRQs with $Y>1$, 700 BL Lacs with $Y\leq1$ and 300 BL Lacs with $Y>1$, are generated that are similar in composition to the historical sample. The physical parameters of these FSRQs and BL Lacs are assigned by generating random numbers from a certain range of values, following either a normal or uniform distribution. It is expected that model generated slopes are consistent with those in Table~\ref{table1}, so we assume that values of $n$ and $x$ of FSRQ and BL Lacs with $Y>1$ respectively conform to normal distributions with a mean of 2.6 and a standard deviation of 0.1, and a mean of 0.3 and a standard deviation of 0.05. For BL Lacs with $Y<1$, values of $x$ conform to a normal distribution with a mean of 2.5 and a standard deviation of 0.1. Values of the other physical parameters conform to uniform distributions within physically plausible ranges. For all blazars, uniform distributions of some parameters have the same range: $0.1^\circ \leqslant \theta_{\rm open} \leqslant3^\circ$ \citep[$\theta_{\rm open}$ represents the jet half opening angle;][]{2019ApJ...870...28F}; $r_{\rm ext} \leqslant \it{r} \leqslant \rm 10^{2}~ pc$; $1 \leqslant \eta \leqslant 10$ \citep{1996ApJ...463..555I, 2001APh....15..121M, 2017ApJ...843..109G}; $1\leqslant \delta_{\rm D,0} \leqslant3$ ($\delta_{\rm D,0}$ represents the initial $\delta_{\rm D}$); $2\leqslant s_{\rm e} \leqslant4$\footnote{Please note that the range of $s_{\rm e}$ here is for the convenience of simulation, and it at odds with the simplest diffusion shock acceleration \citep[e.g.,][]{1998PhRvL..80.3911B}. The value of $s_{\rm e}$ affects the ratio of $L_{\rm inj}(\gamma>\gamma_{\rm c})/L_{\rm rad}$. As mentioned before Eq.~(\ref{FLC}), if $L_{\rm inj}(\gamma>\gamma_{\rm c})$ dominates, all correlations will depend only on if there is a correlation between $\nu$ and $L_{\rm inj}$, which will inevitably change the correlation or increase the dispersion of the correlation largely. When applying the simple scaling model to reproduce the blazar sequence, we only retain blazars with $L_{\rm inj}(\gamma>\gamma_{\rm c})/L_{\rm rad}\leq0.5$ to ensure that the final correlation would not be affected too much. 
Note that, the value of $L_{\rm inj}(\gamma>\gamma_{\rm c})/L_{\rm rad}$ is jointly determined by $s_{\rm e}$ and $\gamma_{\rm c}$. The removal of blazars with $L_{\rm inj}(\gamma>\gamma_{\rm c})/L_{\rm rad}>0.5$ does not contradict the condition $\nu<\nu_{\rm c}$ applied in the simulation.} ($s_{\rm e}$ represents the spectral index of electrons energy distribution). In addition, there are some differences in the range of uniform distributions of $L_{\rm inj}$, $B$ and $L_{\rm D}$. For FSRQs, we set $10^{45}~\rm erg~s^{-1} \leqslant \it{L}_{\rm inj} \leqslant \rm 10^{46}~erg~s^{-1}$; $0.1~\rm G \leqslant \it{B}_{\rm 1pc} \leqslant \rm 1~G$ ($\it{B}_{\rm 1pc}$ represents the magnetic field at 1 pc); $10^{44.5}~\rm erg~s^{-1} \leqslant \it{L}_{\rm D} \leqslant \rm 10^{46.5}~erg~s^{-1}$. For BL Lacs with $Y>1$, we set $10^{43}~\rm erg~s^{-1} \leqslant \it{L}_{\rm inj} \leqslant \rm 10^{44}~erg~s^{-1}$; $0.01~\rm G \leqslant \it{B}_{\rm 1pc} \leqslant \rm 0.1~G$; $10^{42.5}~\rm erg~s^{-1} \leqslant \it{L}_{\rm D} \leqslant \rm 10^{44.5}~erg~s^{-1}$. For BL Lacs with $Y<1$, we set $10^{43}~\rm erg~s^{-1} \leqslant \it{L}_{\rm inj} \leqslant \rm 10^{44}~erg~s^{-1}$; $0.05~\rm G \leqslant \it{B}_{\rm 1pc} \leqslant \rm 0.5~G$.
\begin{figure*}
\centering
\subfloat{
\includegraphics[width=1\columnwidth]{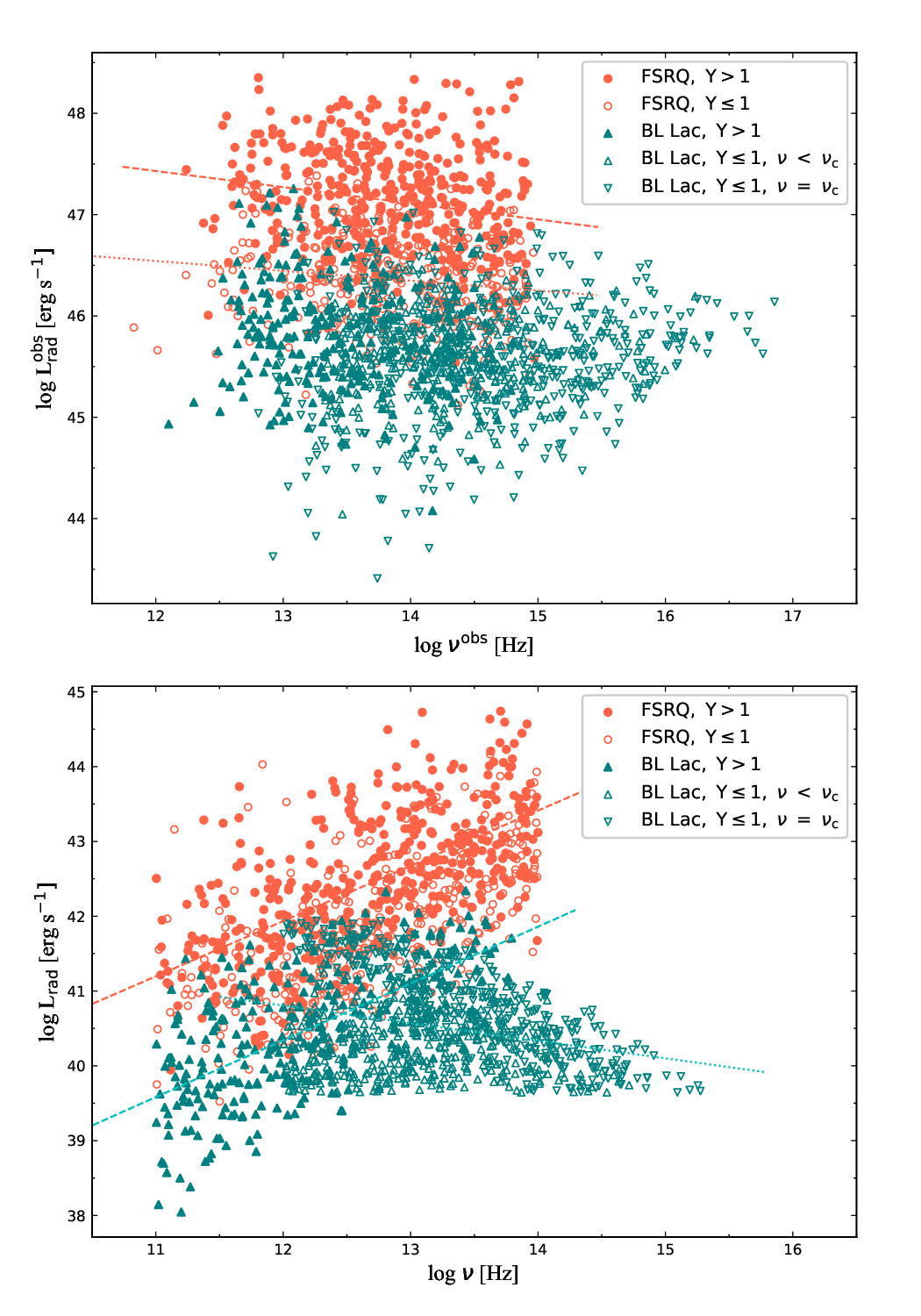}
}\hspace{-5mm}
\subfloat{
\includegraphics[width=1\columnwidth]{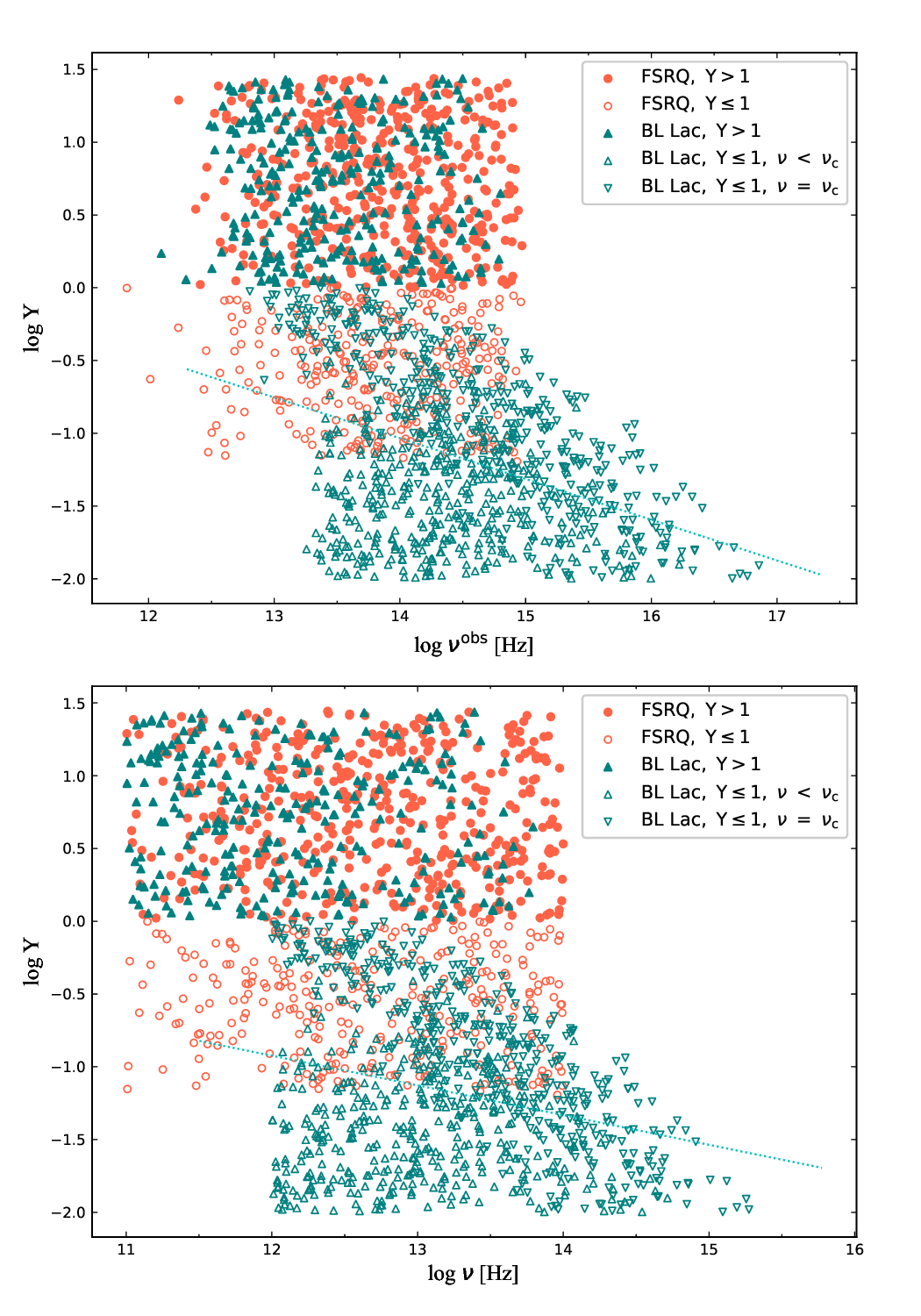}
}
\caption{The phenomenological (upper two panels) and intrinsic (lower two panels) blazar sequence generated by the simple scaling model. Similar to Figures~\ref{correlations_Y}--\ref{correlations_corrected sample}, if a significant correlation is found ($\tau > 0.1$, $p < 0.01$) for a specific subsample statistically, the best linear fitting equation is also shown. Red and teal dashed lines represent the best-fitting equation for the populations of FSRQs and BL Lacs with $Y > 1$, respectively. Red and teal dotted lines represent the best-fitting equation for the populations of FSRQs and BL Lacs with $Y\leq1$, respectively. The meanings of symbols are explained in the inset legends. \label{sim}}
\end{figure*}
Based on the chosen ranges and distributions of parameters described above, the phenomenological and intrinsic blazar sequence generated by the simple scaling model are shown in Figure~\ref{sim}, and the corresponding correlation results are given in Table~\ref{simspear}.  It can be seen that although the model generated correlation results and slopes are basically consistent with those of historical sample and $\delta_{\rm D}$-corrected sample, there are morphological differences between the data distribution obtained from the model and the observed data when viewed with the naked eye. For the correlation results, one major difference is that the slope of the correlation between $\log (L_{{\rm rad}}^{{\rm obs}})$ and $\log (\nu^{{\rm {obs}}})$ for FSRQs with $Y\leq 1$ generated by the model is $-0.10\pm0.04$, while the slope obtained from the historical sample is $-0.35\pm0.08$. This may be due to the fact that the EC emission of these FSRQs with $Y\leq 1$ no longer dominates the high-energy peak, which has changed the slope to some extent. On the other hand, our model, under the current settings, gives the predicted correlation results of FSRQs and BL Lacs with $Y\leq1$, respectively, which can be checked in the future with larger and more complete samples. Distributions of physical parameters are shown in Figure~\ref{para}. It can be seen that parameters distributions are generally reasonable and consistent with those suggested in the conventional leptonic model \citep{2018ApJS..235...39C, 2020Galax...8...72C, 2020ApJS..248...27T}. However, it should be noted that the parameters distributions obtained here are only used to show that the phenomenological and intrinsic blazar sequence can be reproduced by the simple scaling model within a reasonable parameter range, and may not necessarily represent the actual physical properties of blazars. In fact, the blazar sequence currently reproduced using this simple scaling model may not necessarily have universal applicability. Although power-law correlations are built between physical parameters with reasonable assumptions, the normalizations are all random numbers, which greatly increases the dispersion and weakens the correlation. For example, as shown in Table~\ref{simspear}, except for $\log (L_{{\rm rad}})~ {\rm vs.}~ \log (\nu)$, all others are weakly correlated or uncorrelated. If the range of parameter distributions continues to expand, e.g., $L_{\rm inj}$, all the relevant correlations will disappear or the slope will change largely. Actually, \cite{2008MNRAS.387.1669G} suggest that the accretion rate, which is directly related to $L_{\rm inj}$, is important for the explanation of the blazar sequence. Therefore, the model generated results are highly sensitive to the chosen ranges and distributions of parameters. Further study is needed to identify the parameters that truly dominate. In addition, the actual blazar radiation mechanism must be more complex than the model we currently used. For example, the AGN environment contains multiple external photon fields, rather than the current single external photon field assumption; the jet acceleration profile  is more complex and not even continuous; observations indicate that some jets have a parabolic structure \citep{2013ApJ...775..118N, 2017ApJ...834...65A, 2018NatAs...2..472G}. As a result, correlations of the historical sample are mostly weakly correlated or uncorrelated as well, and the correlation obtained from different samples also varies, as discussed in Sect.~\ref{CR}.
\begin{table}
\setlength\tabcolsep{9pt}
\caption{Correlation results of the phenomenological and intrinsic blazar sequence generated by the simple scaling model.}\label{simspear}
\centering
\begin{tabular}{cccc}
\hline\hline														
	&	$r$	&	$p$	&	$s$	\\
FSRQ, $Y>1$	&		&		&		\\
\hline
$ \log (L_{{\rm rad}}^{{\rm obs}})~ {\rm vs.}~ \log (\nu^{{\rm {obs}}})$	&	-0.21	&	$2.5\times10^{-5}$	&	$-0.16 \pm 0.04$	\\
$\log (Y)~ {\rm vs.}~ \log (\nu^{{\rm obs}})$	&	-0.08	&	0.08	&	$-0.05 \pm 0.03$	\\
$ \log (L_{{\rm rad}})~ {\rm vs.}~ \log (\nu)$	&	0.64	&	$2.2\times10^{-53}$	&	$0.74 \pm 0.04$	\\
$\log (Y)~ {\rm vs.}~ \log (\nu)$	&	-0.09	&	0.05	&	$-0.05 \pm 0.02$	\\
\hline							
FSRQ, $Y\leq1$	&		&		&		\\
\hline
$ \log (L_{{\rm rad}}^{{\rm obs}})~ {\rm vs.}~ \log (\nu^{{\rm {obs}}})$	&	-0.15	&	$6.7\times10^{-3}$	&	$-0.10 \pm 0.04$	\\
$\log (Y)~ {\rm vs.}~ \log (\nu^{{\rm obs}})$	&	-0.04	&	0.51	&	$-0.02 \pm 0.03$	\\
$ \log (L_{{\rm rad}})~ {\rm vs.}~ \log (\nu)$	&	0.68	&	$6.6\times10^{-43}$	&	$0.77 \pm 0.05$	\\
$\log (Y)~ {\rm vs.}~ \log (\nu)$	&	-0.07	&	0.22	&	$-0.03 \pm 0.02$	\\
\hline							
BL Lac, $Y>1$	&		&		&		\\
\hline
$ \log (L_{{\rm rad}}^{{\rm obs}})~ {\rm vs.}~ \log (\nu^{{\rm {obs}}})$	&	-0.14	&	0.02	&	$-0.14 \pm 0.05$	\\
$\log (Y)~ {\rm vs.}~ \log (\nu^{{\rm obs}})$	&	-0.08	&	0.16	&	$-0.05 \pm 0.04$	\\
$ \log (L_{{\rm rad}})~ {\rm vs.}~ \log (\nu)$	&	0.64	&	$3.5\times10^{-35}$	&	$0.76 \pm 0.05$	\\
$\log (Y)~ {\rm vs.}~ \log (\nu)$	&	-0.12	&	0.05	&	$-0.07 \pm 0.03$	\\
\hline							
BL Lac, $Y\leq1$	&		&		&		\\
\hline
$ \log (L_{{\rm rad}}^{{\rm obs}})~ {\rm vs.}~ \log (\nu^{{\rm {obs}}})$	&	0.07	&	0.05	&	$0.08 \pm 0.02$	\\
$\log (Y)~ {\rm vs.}~ \log (\nu^{{\rm obs}})$	&	-0.34	&	$4.8\times10^{-24}$	&	$-0.28 \pm 0.02$	\\
$ \log (L_{{\rm rad}})~ {\rm vs.}~ \log (\nu)$	&	-0.22	&	$4.6\times10^{-10}$	&	$-0.24 \pm 0.03$	\\
$\log (Y)~ {\rm vs.}~ \log (\nu)$	&	-0.22	&	$4.6\times10^{-10}$	&	$-0.20 \pm 0.02$	\\
\hline																				
\label{simspear}
\end{tabular}
\end{table}
\section{Conclusion}\label{DC}
In this work, we revisit correlations in the phenomenological and intrinsic blazar sequence across three samples, which are the historical sample, the quasi-simultaneous sample and the $\delta_{\rm D}$-corrected sample, selected from literature. In attempting to interpret these correlations, we propose the simple scaling model, in which physical parameters of the dissipation region are connected to the location of the dissipation region. Our conclusions are as follows:

\textit{Statistical correlation results:} When considering all types of blazars as a whole, the phenomenological blazar sequence holds in the historical sample. In which, a strong negative correlation is found between the Compton dominance and the synchrotron peak frequency, as well as a moderate negative correlations is found between the radiative luminosity and the synchrotron peak frequency. However, the phenomenological blazar sequence does not exist in the quasi-simultaneous sample. It might be caused by the fact that the influence of variability that could cause massive changes in the SED is not considered in \cite{2016MNRAS.463.3038X}. Their main purpose is to make use of the maximum availability of simultaneous data coverage, so SEDs in both low and high states are included. For the intrinsic blazar sequence, correlations between the Compton dominance and the synchrotron peak frequency, and between the radiative luminosity and the synchrotron peak frequency display no correlation and strong positive correlation, respectively. For FSRQs, we find no correlations between the Compton dominance and the synchrotron peak frequency in either observers' or comoving frame. In addition, we find weak negative correlations (excluding the quasi-simultaneous sample) in the observers' frame and a moderate positive correlation in the comoving frame between the radiative luminosity and the synchrotron peak frequency. For BL Lacs with $Y>1$, we find no correlations between the Compton dominance and the synchrotron peak frequency in either observers' or comoving frame, and no correlations in the observers' frame and strong positive correlations in the comoving frame between the radiative luminosity and the synchrotron peak frequency. The derived correlation results and slopes (when correlations exist) are similar to those of FSRQs. For BL Lacs with $Y\le 1$, we find strong negative correlations (excluding the quasi-simultaneous sample) between the Compton dominance and the synchrotron peak frequency, and no correlations between the radiative luminosity and the synchrotron peak frequency in the observers' frame.

\textit{Theoretical implications:} 
In this work, in attempting to reproduce the correlations of the blazar sequence, we propose a simple scaling model. In this model, a dominant dissipation region, which takes into account radiative cooling, is considered to occur along the jet. Consequently, under reasonable assumptions, the physical parameters of the dissipation region are linked to the location of the dissipation region itself. In the modeling, we find that the correlations in the blazar sequence cannot be reproduced satisfactorily unless considering some specific conditions that have been fine-tuned. This implies that radiative cooling alone may not be the primary cause of the blazar sequence. It further indicates that additional physical processes not considered in the simple scaling model is needed to interpret the blazar sequence within a more physically plausible framework. Based on a sensible range of physical parameters, we employ our simple scaling model to generate a population of blazars. The objective is to ascertain whether the observed correlations in the blazar sequence can be accurately replicated with the generated blazar population. Whilst this method is promising to test different hypotheses of the underlying physical mechanism of the blazar sequence, we find that the model generated results are so sensitive to the chosen ranges and distributions of parameters that it may not be able to accurately reproduce the broad properties of the observed blazar population. This demonstrates that this simple scaling model lacks the complexity required to interpret the blazar sequence. Further consideration of black hole mass, accretion rate and a more realistic emission calculation may be required to explain the underlying physics of the blazar sequence.

\section*{Acknowledgements}
We thank the anonymous referees for insightful comments and constructive suggestions. We thank Prof. J. H. Yang for his kindly help and sharing data. This work is supported by the National Natural Science Foundation of China (NSFC) under the Grants No. 12203043. Z.R.W. acknowledges the support by the NSFC under the the Grants No. 12203024. H.B.X. acknowledges the support by the NSFC under the the Grants No. 12203034 and by Shanghai Science and Technology Fund under the Grants No. 22YF1431500. J.H.F. acknowledges the support by the NSFC under the the Grants No. U2031201.

\section*{Data Availability}
The data underlying this article will be shared on reasonable request to the corresponding author.



\bibliographystyle{mnras}
\bibliography{example} 








\bsp	
\label{lastpage}
\end{document}